\documentclass[aps,pre,twocolumn,groupedaddress]{revtex4}

\usepackage{amsmath}
\usepackage{amsfonts}
\usepackage{amssymb}
\usepackage{bm}
\usepackage{color}
\usepackage{graphicx}
\begin{document}

\title{Stochastic spatial models in ecology: a statistical physics approach}
\author{Simone Pigolotti$^1$
Massimo Cencini$^2$,
 Daniel Molina$^3$
Miguel A. Mu\~{n}oz$^4$
}

\affiliation{$^1$ Biological Complexity Unit, Okinawa Institute of
  Science and Technology and Graduate University, Onna, Okinawa
  904-0495. $^2$ Istituto dei Sistemi Complessi, Consiglio Nazionale
  delle Ricerche, via dei Taurini 19, 00185 Rome, Italy.  $^3$ BCAM -
  Basque Center for Applied Mathematics, Alameda de Mazarredo 14,
  E-48009 Bilbao, Basque Country, Spain.  $^4$ Departamento de
  Electromagnetismo y F\'{i}sica de la Materia, and Instituto Carlos I
  de F\'{i}sica Te\'{o}rica y Computacional, Universidad de Granada,
  18071 Granada, Spain.  }

\begin{abstract}

  Ecosystems display a complex spatial organization. Ecologists have
  long tried to characterize them by looking at how different measures
  of biodiversity change across spatial scales. Ecological neutral
  theory has provided simple predictions accounting for general
  empirical patterns in communities of competing species. However,
  while neutral theory in well-mixed ecosystems is mathematically well
  understood, spatial models still present several open problems,
  limiting the quantitative understanding of spatial biodiversity.  In
  this review, we discuss the state of the art in spatial neutral
  theory. We emphasize the connection between spatial ecological
  models and the physics of non-equilibrium phase transitions and
  how concepts developed in statistical physics translate in
  population dynamics, and vice versa. We focus on non-trivial scaling
  laws arising at the critical dimension $D = 2$ of spatial neutral
  models, and their relevance for biological populations inhabiting
  two-dimensional environments. We conclude by discussing models
  incorporating non-neutral effects in the form of spatial and
  temporal disorder, and analyze how their predictions deviate from
  those of purely neutral theories.

\end{abstract}

\maketitle

\section{Introduction}
\label{sec:intro}

Community ecology aims at shedding light on how competing species
assemble and coexist in their habitats \cite{community}. This
has proven to be a formidable challenge.  A main reason is that
ecological dynamics span a wide range of spatial and temporal scales,
from those typical of individuals to those characterizing large
populations or communities. Ecologists have
empirically characterized biodiversity at the different spatial
scales; for example, counting the average number of species hosted in
a given area -- {\em species area relationship} (SAR)
\cite{preston62,rosenzweig1995}--, or the distribution of their
abundances -- {\em species abundance distribution} (SAD)
\cite{macarthur1960relative,tokeshi1993species}. Often, the ecological
forces determining these patterns act at a given spatio-temporal
scale but can affect others as well.  The inverse problem,
i.e. linking observed patterns with the causes originating
them at different scales, is arguably the central problem in ecology
\cite{levin1992problem}.

This kind of problem sounds familiar to experts in statistical
physics, where large-scale emergent behavior results from interactions
among simple local units. Tools of statistical physics are indeed very
useful to make progress on the aforementioned crucial issues in
ecology. In particular, a natural approach to such complex problems is
to radically simplify them. To this aim, we consider ecosystems made
up of competing non-motile species, such as trees, or having a
motility range much smaller than the typical linear size of the
population, such as communities of
microorganisms. Further possible simplifications are
that all emergent phenomena originate at the single-individual scale
and, more drastically, that differences among individuals, possibly
belonging to different species, can be neglected. These assumptions
constitute the basis of the {\em ecological neutral theory} proposed
by Hubbell \cite{Hubbell2001}.

Ecological neutral theory \cite{Hubbell2001} was built upon
theoretical ideas of Kimura's neutral theory of population genetics
\cite{kimura1983neutral}. Both theories underscore the
role of stochastic demographic fluctuations in determining the fate of
populations and completely neglect deterministic effects stemming
  from fitness differences.  The assumption of ecological neutrality has elicited
heated controversies, as it hinted that classical ecological
  concepts, such as niches, might play a marginal role in
structuring communities of competing species.  Despite these
contentions, neutral theory had a considerable impact on
ecological thinking, owing to its ability to quantitatively predict
non-trivial patterns of biodiversity with simple models
characterized by very few
adjustable parameters
\cite{alonso2006merits,rosindell2011unified,azaele2016statistical}.

Spatially implicit neutral models describe well-mixed communities of
individuals subject to immigration from a larger reservoir of species
where diversity is maintained via speciation. They can be solved
analytically
\cite{vallade2003analytical,mckane2004analytic,Volkov2003,pigolotti2005species,etienne2007neutral},
yielding analytical expressions for the SAD. Beside the mathematical
appeal, these exact solutions have been extremely helpful for fitting
empirical data and therefore testing neutral theory or, at least,
promote it as a null-model \cite{rosindell2012case}.  For more
  exhaustive surveys of ecological neutral theory, we refer the
reader to Hubbell's book \cite{Hubbell2001} and the reviews
\cite{alonso2006merits,rosindell2011unified,azaele2016statistical}.

The focus of this review  is on spatially-explicit neutral and
near-neutral population models. Explicitly describing space is crucial
to address the fundamental ecological questions sketched at the
beginning of the introduction.  However, spatially-explicit models --
that are often
variants of familiar models in non-equilibrium statistical physics
\cite{durrett1994stochastic} --  are
still poorly understood, especially if compared with their well-mixed
counterparts \cite{etienne2011spatial}.  One of the most studied
neutral model is the {\em voter model with speciation}, or
multi-species voter model
\cite{durrett1996spatial,Rosindell2007,Pigolotti2009}, which
generalizes the more common two-species voter model
\cite{Liggett1985}. The stepping-stone model
\cite{kimura1953,kimura1964stepping,korolev2010genetic,cencini2012ecological}
and the contact process \cite{Liggett1985,MarroBook,Hinrichsen2000}
are other examples of spatial models that have been studied in
both the physics and population biology literature. We shall
discuss how these analogies can be used to advance our understanding
of spatial ecology and the main open problems.  This review heavily
relies on extensive numerical computations of lattice models based on
previous works by the authors. This might have biased the choice of
some topics and we apologize if some relevant works are not properly
discussed.

The review is organized as follows. In Sect.~\ref{sec:vm} we introduce
the multispecies voter model on a lattice and its dual representation
in terms of coalescing random walkers. We then discuss its predictions
of macroecological patterns: the SAR, and the SAD. For the latter, we
compare two recent analytical approaches
\cite{PNAS-Zillio,Danino2,danino2016spatial} with novel computational
results.  We mainly discuss the two-dimensional case
  due to its ecological relevance, but also briefly present the
  one-dimensional case for comparison. We conclude the section by
  presenting new results on an important dynamical property: the distribution of
  species persistence-times.  In Sect.~\ref{sec:otherneutral} we
discuss other neutral models, where, at variance with
the voter model, lattice sites are not necessarily occupied by exactly
one individual at all times. In particular, we consider the stepping
stone model, where each lattice site hosts a local community of
individuals. This generalization is relevant for
modeling microorganisms and their macroecological patterns. We then
consider a multispecies variant of the contact process, where lattice
sites can be either empty of occupied by single individual.  In
Section~\ref{sec:nearneutral} we introduce non-neutral effects on a
simplified two-species competition model, where adjusting a single
parameter one can tune the departure from neutrality, here modeled as
a specific habitat preference. Physically, this habitat preference can
be thought as a form of quenched disorder. We discuss how this
disorder generically favors species coexistence using the language of
statistical mechanics, and also discuss other forms of disorder such
as temporal heterogeneity. Finally, Sect.~\ref{sec:conclusions} is
devoted to perspectives and conclusions.

\section{Voter Model with speciation\label{sec:vm}}

\subsection{Description of the model}
\label{sec:2}

A paradigmatic example of spatial neutral model is the {\em voter
  model with speciation}, \cite{durrett1996spatial}, which is is a
multispecies generalization of the voter model
\cite{Liggett1985}. The latter is a widely studied model
that has been applied in diverse contexts, from
population genetics to spatial conflicts \cite{clifford}, spreading of
epidemic diseases \cite{pinto}, opinion dynamics
\cite{reviewCastellano} and linguistics \cite{croft}.

\begin{figure}[htb]
\begin{center}
\includegraphics[width=8cm]{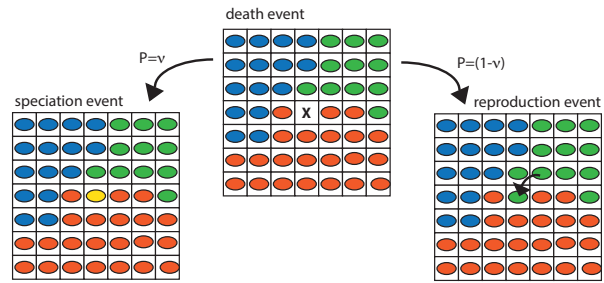}
\caption{Examples of transitions in the $2D$ voter model
  with speciation. \label{fig:vm} }
\end{center}
\end{figure}

The voter model with speciation is defined on a lattice, where each
site hosts one individual belonging to some species.  At each discrete
time step, a lattice site is chosen at random and the residing
individual is removed (death event). Then,
as illustrated Fig.~\ref{fig:vm}, the dead individual is replaced:
\begin{itemize}
\item With probability $\nu$, by an
  individual of a new species not present in the system (speciation
  event). Notice that, because of speciation, the total number of
  species is not fixed.  In population genetics, this type of event is
  interpreted as a mutation within the same species
  \cite{Moran58,kimura1964stepping}.
\item With complementary probability $(1-\nu)$, by a new individual of
  an existing species (reproduction event).  In this case, the newborn
  belongs to the same species of a parent individual
  chosen at random in the neighborhood of the vacant site. In the
  simplest case, the nearest-neighbors (NN) are chosen with uniform
  probability. More generally, the parent individual is selected
  according to a probability distribution $P(\vec{r})$ (the {\em
    dispersal kernel}) over the neighbors within a
  distance $\vec{r}$.
\end{itemize}
Most of this section will be devoted to the ecologically relevant case
where the system is a two-dimensional ($2D$) square lattice, 
although we will briefly present some results in $1D$
for comparison.

\subsection{Duality\label{sec:duality}}

The voter model with speciation is {\em dual} to a system of
coalescing random walkers with an annihilation rate
\cite{Liggett1975,Bramson1996,durrett1996spatial}.  In this
  context, ``duality'' means that each trajectory of one system can be
  mapped in one of the other system having equal probability
  \cite{Liggett1975}. The dual process is constructed as follows.  We start by placing on each lattice
site a random walker.  The dynamic of the dual process proceeds
backward in time. At each discrete (backward) time step, with
probability $1-\nu$, a randomly chosen walker is moved to a new site,
where the dispersal kernel $P(\vec{r})$ here plays the role of the
distribution of possible displacements. If the site is occupied, the
two walkers coalesce, i.e. one of the two is removed keeping trace of
the coalescing partner.  With complementary probability $\nu$ a
randomly chosen random walker is annihilated, i.e. removed from the
system. This event corresponds to a speciation event in the forward
dynamics.  The whole tree of coalescing random walkers, before
annihilation, represents the entire genealogical tree of a species up
to the speciation event that originated it.

\begin{figure}[thb]
\begin{center}
\includegraphics[width=8cm]{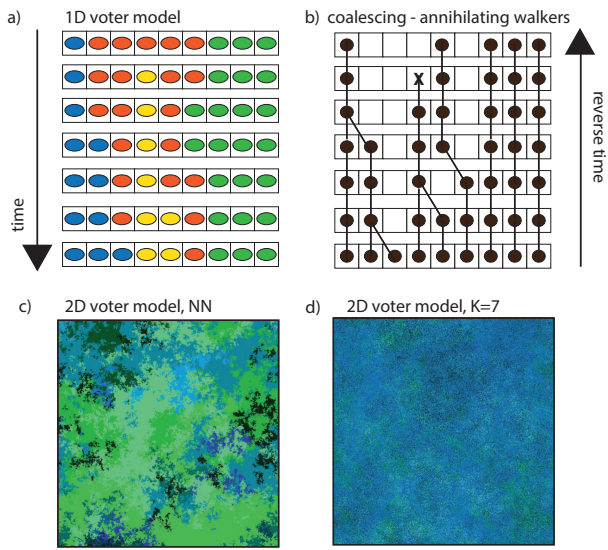}
\caption{a) Example of space-time dynamics of the $1D$ voter model
  with speciation. b) Corresponding dual dynamics: coalescing and
    annihilating random walkers. c) Snapshot of a configuration of
  the $2D$ voter model simulated with the dual dynamics, with
  $\nu=5\ 10^{-7}$ and nearest-neighbor (NN) dispersal. d) Same as c)
  but with a longer dispersal range (uniformly distributed in a square of
  side $K$) with $K=7$. Each color labels a different
  species. \label{fig:dual}}
\end{center}
\end{figure}

The standard forward in time evolution of the voter-model with
speciation and its dual dynamics are sketched, for
the one-dimensional case, in Fig.~\ref{fig:dual}a and \ref{fig:dual}b,
respectively.

Duality is a very useful property to understand the physics of the
voter model. For example, it immediately stems from duality that the
$\nu\rightarrow 0$ limit is fundamentally different in $D\leq 2$ and
$D>2$. As a matter of fact, in $D\leq 2$ the random walk is recurrent,
meaning that the probability of two randomly chosen individuals to
belong to the same species approaches one as $\nu\rightarrow 0$. In
other words, in the absence of speciation, one has
monodominance of one species in the long term. The same property does
not hold in $D>2$, where random walkers are not recurrent and, 
in an infinite system, multiple species coexist on the long term even in the
limit $\nu\to 0$. Interestingly, the ecologically most relevant case,
$D=2$, is the critical dimension of this model. We shall see that this
fact is a source of non-trivial behaviors of
ecologically relevant quantities.

Duality is also an extremely powerful tool for computational analyses
\cite{Rosindell2007,Pigolotti2009}. If one is interested in the
static, long-term, properties of the voter model with speciation, it
is numerically much more efficient to simulate the dual dynamics than
the forward one. In a dual simulation, after all walkers coalesced or
were annihilated, species can be assigned to the start site of each
walker, obtaining a stationary configuration of the voter
model. Beside computational speed, this approach has also the
advantage of eliminating finite-size effects induced by the boundary
conditions, as the coalescing random walkers can be simulated in a
virtually infinite system. For illustrative purposes, in
Fig.~\ref{fig:dual}c and \ref{fig:dual}d we show two configurations of
the $2D$ voter model obtained with the dual dynamics for two different
dispersal kernels.

%%%%%%%%%%%%%%%%%%%%%%%%%%%%%%%%%%%%%%%%%%%%%%%%
\subsection{ $\beta-$diversity\label{sec:betadiversity}}
%%%%%%%%%%%%%%%%%%%%%%%%%%%%%%%%%%%%%%%%%%%%%%%%  

The first ecological pattern we consider is the 
$\beta$-diversity, which is a measure of how the species composition
in an ecosystem varies with the distance. We define the
$\beta$-diversity as the probability $F(\vec{r})$, that two
randomly chosen individuals at a distance $\vec{r}$ are conspecific,
i.e. belong to the same species. We remark that,
  although this is the natural definition in this context, other
  definitions have been used
in the ecological literature \cite{beta}.  Mathematically, $F(\vec{r})$
can be expressed in terms of the two-point correlation function
$G_{s_i,s_j}(\vec{r})=\langle
n_{s_i}(\vec{x})n_{s_j}(\vec{x}+\vec{r})\rangle$, where
$n_{s_i}(\vec{x})$ denotes the number of individuals of species $s_i$
at location $\vec{x}$
\begin{equation}
F(\vec{r}) = \frac{\sum_i G_{s_i,s_i}(\vec{r})}{\sum_{i,j} G_{s_i,s_j}(\vec{r})}\,,
\label{eq:betacorr}
\end{equation}
where the sums extend over all species in the ecosystem
\cite{azaele2016statistical}. Eq.~(\ref{eq:betacorr}) can be used
  to estimate the $\beta$-diversity as the ratio between the
couples of conspecific over the total number of couples in a sample.

Let us now study the evolution equation of $F(\vec{r},t)$ for the
voter model with speciation and 
NN  dispersal. Although we shall focus on the $2D$ case,
 it is useful to present the general calculation
in $D$ dimensions.  Following
\cite{chave2002spatially,PRL-Zillio,azaele2016statistical} we write
\begin{eqnarray}
  F(\vec{r},t+1)&=&\left(1-\frac{2}{N} \right)F(\vec{r},t)+\\
&+&\frac{1-\nu}{DN}\sum_{k=1}^{D}[F(\vec{r}+\vec{e}_k,t)+F(\vec{r}-\vec{e}_k,t)]\,. \nonumber
\label{eq:beta1}
\end{eqnarray}
The first term in the r.h.s. of Eq. (\ref{eq:beta1}) represents the
fact that $F$ does not change if two generic individuals at distance
$\vec{r}$  are not removed in a given time step and
  therefore survive. The second term represents the events in which
one of the two individuals dies (with prob. $2/N$), no speciation
occurs (with prob. $1-\nu$) and the dead individual is replaced by a
conspecific from the $2D$ neighbor sites.  Taking the continuous limit
$N\to \infty$ with the lattice spacing $a\to 0$, the speciation
probability $\nu\to 0$, and a finite value of $\kappa^2=2D\nu/a^2$, one obtains
at stationarity the differential equation
\begin{equation}
  \frac{1}{r^{D-1}}\frac{d}{dr}r^{D-1}\frac{dF}{dr}-\kappa^2 F(r)+ c
  \delta^{D}(r)=0
  \label{eq:beta2}
\end{equation}
where $\delta^D$ is the $D-$dimensional Dirac delta, and because of
isotropy the $\beta$-diversity $F(r)$ is now function of $r=|\vec{r}|$
only. The solution of
Eq.(\ref{eq:beta2}) is \cite{azaele2016statistical}
\begin{equation}
  F(r)= c\frac{\kappa^{D-2}}{(2\pi)^{D/2}} (\kappa r)^{(2-D)/2} K_{(2-D)/2}(\kappa r)\,,
  \label{eq:beta3}
\end{equation}
where $K_{z}$ is the modified Bessel function of the second kind of
order $z$ and the constant $c$ is fixed by the condition
$\int_{r<a} d^Dr F(\vec{r})=1$. We recall that Eq.~(\ref{eq:beta3}) is a
continuous expression, valid for distances much larger than the
lattice spacing \cite{chave2002spatially}. Although we derived
Eq.~(\ref{eq:beta3}) for NN dispersal, the same results
hold for a general dispersal kernel for distances larger than the
kernel range, provided that the kernel range is finite.

For $D=2$, Eq.~(\ref{eq:beta3}) implies that $F(r) \propto K_0(\kappa
r)$, which is characterized by a slow logarithmic decay, $\sim
-\ln(r\kappa)$, up to distances of order $1/\kappa \sim 1/\sqrt{\nu}$,
followed by a faster, exponential falloff.  Remarkably, the
$\beta$-diversity empirically measured in several tropical forests in
Central and South America is consistent with a logarithmic decay for
large distances \cite{condit2002beta}.  We remark that this
logarithmic decay is the signature that $D=2$ is the critical
dimension for the voter model. In contrast, in $D=1$,
Eq.~(\ref{eq:beta3}) becomes $F(r) \propto
\sqrt{r\kappa}K_{1/2}(\kappa r) \sim \exp(-r\kappa)$.
We mention for later convenience that, in $D=1$ with
  NN dispersal, Eq. (\ref{eq:beta1}) can be solved 
  without using the continuous approximation,
  giving \cite{PRL-Zillio}
\begin{equation}
F(r)= \exp(-\alpha(\nu) r)\,,\ \ \mathrm{with}\ \ \alpha(\nu)=\ln\left[{\scriptstyle \frac{(1-\nu)}{(1-\sqrt{(\nu(2-\nu)})}}\right]\,,
\label{eq:2punti}
\end{equation}
where $\alpha(\nu) \approx \sqrt{(2\nu)}$ for $\nu\to 0$.

Although the $\beta$-diversity decays exponentially on scales
$1/\kappa \sim 1/\sqrt{\nu}$ both in $1D$ and $2D$, there are
important differences. Because $2D$ is the critical dimension, a large
biodiversity (i.e. a large average number of species) can be sustained
by very low values of the speciation rate $\nu$. This implies that in
$2D$ there are many species living on scales much smaller than
$1/\kappa$, where the correlations decay logarithmically. Conversely,
in $1D$ to maintain biodiversity one needs a large value of $\nu$, so
that $1/\kappa$ is the only characteristic scale and there is no
additional structure on scales smaller than $1/\kappa$. This
crucial point will be further elucidated in the rest of the section,
where we will discuss other observables in $2D$ (subsections
\ref{sec:VMSAR} and \ref{sec:sad}) and compare them with their
  $1D$ counterparts.

%%%%%%%%%%%%%%%%%%%%%%%%%%%%%%%%%%%%%%%%%%%%%%%%
\subsection{Species-Area Relationships\label{sec:VMSAR}}
%%%%%%%%%%%%%%%%%%%%%%%%%%%%%%%%%%%%%%%%%%%%%%%%

We now focus on the SAR, defined as the average number of species, $S$
of a given taxonomic level occupying a given area of size $A$. SARs
are widely studied as a measure of spatial biodiversity
and quantify how larger habitats support more species than
smaller ones \cite{rosenzweig1995}.  Empirical measures of SARs at
multiple scales often reveal three different regimes
\cite{preston62,rosenzweig1995,Hubbell2001}. At small areas, the
number of species increases rather steeply, nearly linearly, with the
sampled areas. A similar steep increase is observed at very large,
continental scales. Instead, at intermediate scales, a slower,
sublinear growth is often found. Such a growth is well
approximated by a power law $S\sim A^z$, $z<1$, over a wide range of
taxa \cite{arrhenius1921}, though a logarithmic behavior
$S \approx C\ln A$ has also been proposed.  An extensive meta-study by
Drakare et al. \cite{drakare2006imprint} reconsidered a large body of
SAR studies from the literature, revealing that the power law provides
a better fit in about half of the cases.  This study
also observed that the exponent $z$ correlates positively with the
body size of the considered group of species, so that small
microorganisms typically display very shallow SAR curves as compared
with larger organisms (see also \cite{horner2004} and
Sect.~\ref{sec:ssm}).

\begin{figure}[ht!]
  \includegraphics[width=8cm]{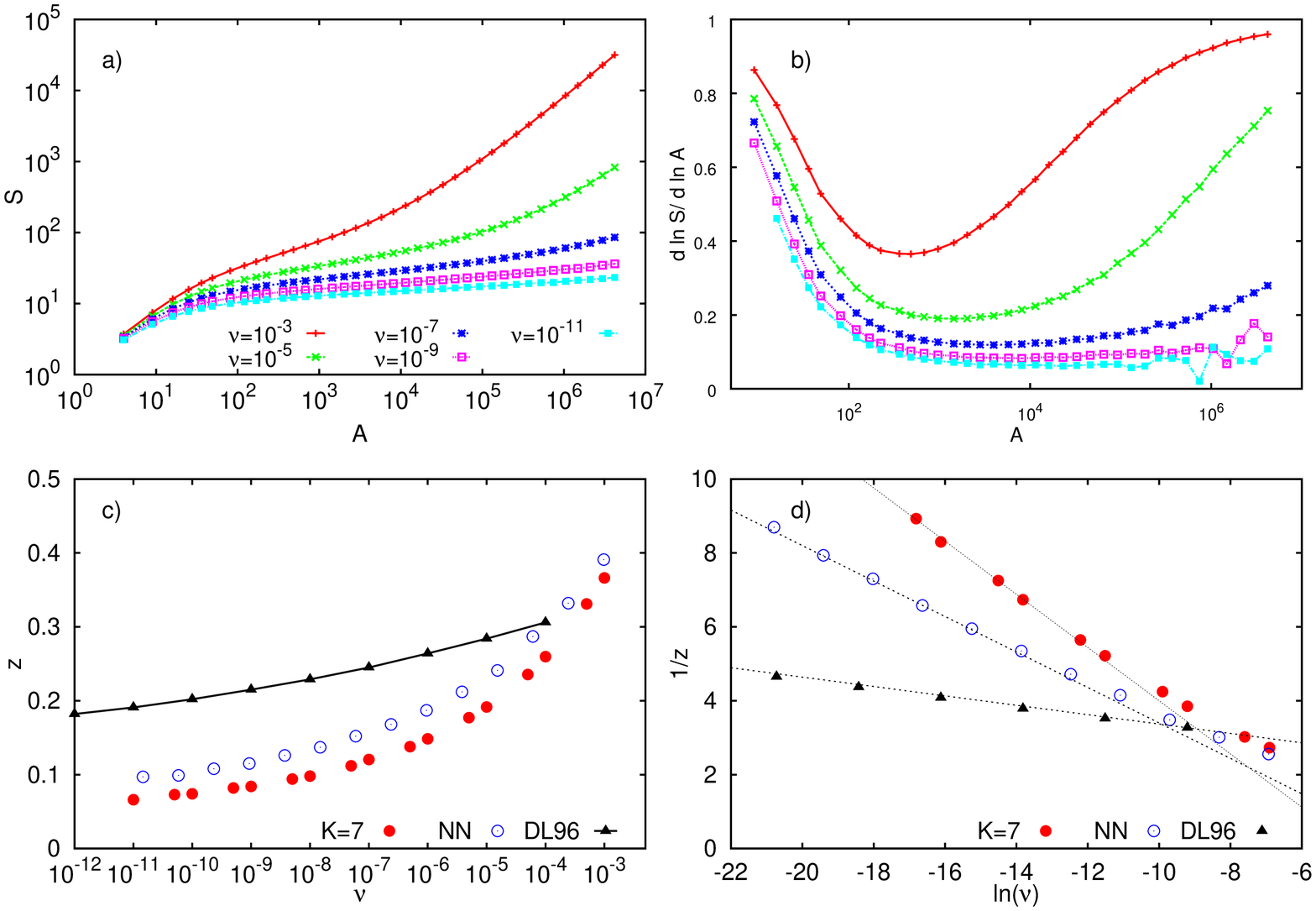}
  \caption{Species Area Relationships (SAR) and their scaling behavior
    in the voter model with speciation. a) Number of species $S$ as a
    function of the sampled area $A$ for different speciation rates as
    in the caption. The triphasic shape is evident for larger
    speciations rates. Simulations from \cite{Pigolotti2009} were
    performed with a square dispersal kernel, i.e. $P(\vec{r})$ is a
    uniform distribution on a square of side $K$ centered on the empty
    site, with $K=7$. b) Local slopes, $d\ln S/d\ln A$ for the curves
    shown in panel a. c) Dependence of the exponent $z$ on $\nu$ as
    obtained from the local slopes for both the square kernel with
    $K=7$ and NN dispersal. The exponent is estimated at the
    inflection point of the SADs, i.e. at the minimum of the local
    slopes. Also shown is the prediction of Eq.(\ref{eq:dl}) (black
    solid line) where the black triangles correspond to the values
    provided in \cite{durrett1996spatial}. d) Plot of $1/z$ vs
    $\ln(\nu)$ of the same data of panel c to highlight the
    logarithmic behavior of Eq.~(\ref{eq:fitlog}) \label{fig:vm_sad}}
\end{figure}

Simulations of the (dual) voter model with speciation yields SARs
qualitatively similar to those obtained from field data, see
Fig.~\ref{fig:vm_sad}a. In the voter model, the steep initial regime
is mostly determined by the dispersal range $K$. For areas
significantly larger than $K^2$, a sublinear growth is observed (see
Fig.~\ref{fig:vm_sad}b.  In this regime, the growth
becomes progressively more shallow as the speciation rate $\nu$ is
decreased. For larger scales, the logarithmic slope
of the SAR curves become steeper again. The area at which this final
crossover occurs increases as $\nu$ is decreased.

An interesting question is whether the sublinear growth regime in the
voter model can be characterized by a power-law $S\sim A^z$ and, in
this case, what is the value of the exponent $z$ as a function of
  $\nu$. To address this question, we begin by
  reviewing a classic estimate of $z$ by Durrett and Levin
\cite{durrett1996spatial} relying on duality (see
Sect.~\ref{sec:duality}). The speciation rate $\nu$ sets a time scale
$1/\nu$ which also corresponds to a characteristic length scale
$\xi=1/\sqrt{\nu}$ because of the diffusive behavior of random walkers
in the dual model.  Walkers with an initial separation much
larger than $\xi$ are likely to be annihilated before
coalescence occurs. This observation alone explains the linear scaling
of $S(A)$ for areas $A\gg \xi^2=\nu^{-1}$. At these scales, species
are uncorrelated, as can also be inferred from the analysis of the
$\beta$-diversity in the previous section.  For a system of coalescing
random walkers in $2D$, the density of occupied sites $\rho(t)$ decays
asymptotically as \cite{bramson1991,MF}
\begin{equation}\label{bram}
\rho(t)\sim \frac{\ln t}{\pi t} \, .
\end{equation}
The characteristic logarithmic coarsening of clusters observed in the $2D$
voter model without speciation can be related to the logarithm
appearing in Eq.~(\ref{bram}) \cite{dornic2001critical}.  Assuming
$\nu\ll 1$, the annihilation rate at time $t$ in an area
$\xi \times \xi$ can be approximated as the annihilation rate per
walker $\nu$ times the number of walkers in the absence of
annihilations $\xi^2\rho(t)$. Integrating over time, we find that the
total number of annihilations, i.e. the total number of species, is
\cite{Bramson1996}
\begin{eqnarray}\label{eq:sxi_vm}
S(\xi^2) &\approx& \nu\xi^2\int_{t_0}^{1/\nu=\xi^2} dt \ \rho(t) =
\frac{\ln^2(\xi^2)-\ln^2(t_0)}{2\pi}\approx\nonumber\\
&\approx &
 \frac{2}{\pi}\left(\ln \xi\right)^2 \,,
\end{eqnarray}
where $t_0$ is the time at which the asymptotic expression
(\ref{bram}) starts to be valid.  The upper temporal cut-off is set to
$1/\nu$ (with $1/\nu=\xi^2$) because the number of killing events
occurring after a time $\sim 1/\nu$ is bounded by the number of
walkers in the system, which is $\xi^2\rho(1/\nu) \sim \ln \xi$
\cite{Bramson1996}.  Finally, combining Eq.~(\ref{eq:sxi_vm}), the
fact that $S(1)=1$ and matching a power law behavior $S=A^z$ in the
range of scales from $A=1$ to $A=\xi^2$, one finds
\cite{durrett1996spatial}
\begin{equation}\label{eq:dl}
z=\frac{\ln \left[S\left(A\right)\right]}{\ln 
  \left(A\right)}=\frac{2\ln [\ln (1/\sqrt{\nu})]+\ln(2/\pi)}{\ln (1/\nu)}\, . 
\end{equation}
Also in this case, the logarithmic dependence of the exponent $z$
on $\nu$ derives from the fact that $D=2$ is the critical dimension for
the voter model.

More recent results disputed the validity of Eq.~(\ref{eq:dl}).
Scaling arguments hinted that $z$ should approach a finite value
$z\approx 0.2$ in the limit of vanishing $\nu$ (see \cite{PRL-Zillio}
and Sec. \ref{subsub:naive}), while numerical simulations suggested a
power law dependence, $z\sim \nu^{0.15}$
\cite{Rosindell2007}. Finally, further numerical simulations, based on
the dual representation of the voter model with speciation (see
Sect. \ref{sec:duality}) and spanning a very wide range of speciation
rates from $10^{-3}$ to $10^{-11}$ confirmed the logarithmic behavior
predicted by Eq.~(\ref{eq:dl}) \cite{Pigolotti2009}. The exponents
measured in such simulations, shown in Fig.~\ref{fig:vm_sad}c, are
well fitted by a phenomenological expression of the form
\begin{equation}
  z=\frac{1}{q+m\ln(\nu)}
  \label{eq:fitlog}
\end{equation}
which is consistent with Eq.~(\ref{eq:dl}) up to order $\ln \ln \nu$,
see also Fig.~\ref{fig:vm_sad}d. However, fitted values of the prefactors
$q$ and $m$ are not consistent with Eq.~(\ref{eq:dl}). This discrepancy is
probably due to pre-asymptotic effects as well
  as to the approximation of assuming a power-law range between $A=1$
and $A=\ln(1/\nu)$.

Let us briefly discuss the role of
the dispersal kernel. As illustrated in Figs.~\ref{fig:vm_sad}c
  and \ref{fig:vm_sad}d, a comparison between NN dispersal and a
square dispersal kernel of range $K=7$ demonstrates that the exponent
$z$ depends to some extent on the dispersal kernel.  However,
numerical evidence \cite{Rosindell2007,Pigolotti2009} suggests that
when the dispersal kernel range is large enough (approximately
$K\geq 5$) the exponents are very weakly dependent on $K$. Moreover,
SARs obtained with different values of $K$ can be rescaled onto a
universal function of $A$ and $\nu$ via the 
transformation $S=f(A,\nu,K)=K^\chi\phi(A/K^\chi,\nu)$ with a fitted
value of $\chi\approx 1.97$. To the best of our knowledge, a formal
derivation of this scaling law and of the exponent $\chi$ is currently
an open problem.  

The non-trivial area dependence of the SAR results is
  a special feature of the critical dimension $D=2$. To highlight this
  point, we now discuss the $D=1$ case as comparison. This
  case is also relevant to describe quasi one-dimensional ecosystems,
  such as river basins \cite{suweis2012}.  For simplicity, we limit
  ourselves to the case of NN dispersal.

\begin{figure}[t!]
\begin{center}
\includegraphics[width=9cm]{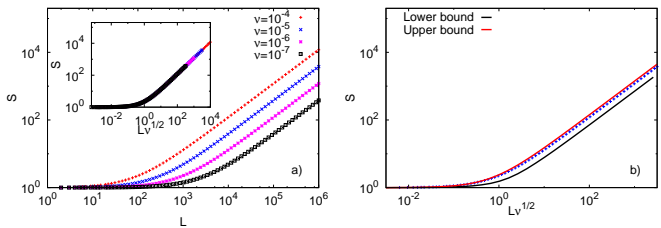} 
\caption{Species Area Relationshipt for the voter model in $D=1$. a)
  Average number of species $S$ versus the system size $L$ for
  different $\nu$ as labeled. Inset: same curves plotted vs
  $L\sqrt{\nu}$, notice the excellent collapse. b) SAR for
  $\nu=10^{-5}$ compared with the theoretical upper
  (\ref{eq:upper}) and lower (\ref{eq:lower}) bounds.
\label{fig:1D}}
\end{center}
\end{figure}

To the best of our knowledge, also in $D=1$, an exact
  expression for the average number of species, $S(L)$, in a segment
  of length $L$ is unknown. Nevertheless, it is possible to provide a
  lower and upper bound for $S(L)$.  In $D=1$, the density of walkers
  behaves as $\rho(t) \sim 1/\sqrt{t}$, to be contrasted with
  eq. (\ref{bram}) valid in the $2D$
  case. Dimensional arguments then suggest that the
  average number of species must a function of $L\sqrt{\nu}$ only,
  i.e. $S(L;\nu)=\Psi(L\sqrt{\nu})$.  Computational results
  (Fig.~\ref{fig:1D}a and inset) support well this simple argument. As
  shown in the figure, the non-trivial power-law regime characteristic
  of $2D$ SARs is absent in $D=1$. Indeed, the function $\Psi$ is
  linear for large arguments, with a coefficient around $1.2$ and it
  is nearly constant for $L\sqrt{\nu} \ll 1$.

We can derive an upper bound to $S(L)$ using that,
  in $D=1$, individuals are organized in $N_s(L;\nu)$ segments of
  conspecific individuals, so that $S\leq N_s$, with the
  equality holding if no species is present in more than one segment.
  We compute $N_s$ from the probability $P_{i-1,i}\equiv F(|i-j|)$,
  with $F(r)$ given by Eq.~(\ref{eq:2punti}),  that two sites $i$
  and $j$ are occupied by conspecific individuals
  \cite{derrida1999genealogical}
\begin{eqnarray} 
S\leq N_s&=&L-\sum_{i=1}^{L-1} P_{i-1,i}= L-(L-1)F(1)=\nonumber\\
&=& L-(L-1)e^{-\alpha(\nu)}  \,,
\label{eq:upper}
\end{eqnarray}
which for $\nu\to0$ can be approximated as $N_s\approx 1 + \sqrt{2\nu}(L-1)$.

The lower bound follows from Jensen's
  inequality (see also \cite{derrida1999genealogical})
applied to the frequency of species represented by the individual in
site $i\in [0,L-1]$, here denoted $n_i(L)$, which yields
\begin{equation}
S=\sum_i \left\langle\frac{1}{n_i(L)}\right\rangle 
\ge \sum_i \frac{1}{\langle n_i(L)\rangle}\,,
\label{eq:lower}
\end{equation}
where $\langle n_i\rangle=\sum_j P_{i,j}$ and $P_{i,j}=F(|i-j|)$
is again given by Eq.~(\ref{eq:2punti}) and can be easily summed
numerically.

In Fig.~\ref{fig:1D}b we compare the numerically
  obtained SAR with the upper (\ref{eq:upper}) and lower 
  (\ref{eq:lower}) bounds. Notice that the upper bound is very close to the
  actual SAR, implying that most species are organized in single
  segments.

\subsection{Species-Abundance Distributions}\label{sec:sad}

We now discuss Species-Abundance Distributions (SADs), $P(n;A)$, that
measure the relative abundance of species in a given area $A$. More
precisely, denoting $S(A)$ the total number of species sampled in an
area $A$, each composed by $n_i$ ($i=1,\ldots,S(A)$) individuals,
$P(n;A)dn$ is the probability that a randomly picked species has an
abundance between $n$ and $n+dn$. While the expression of $P(n;A)$ for
well-mixed neutral models is known \cite{Volkov2003}, computing it for
spatially explicit models, such as the voter model with speciation,
has proven to be a rather hard problem.  We first discuss in section
\ref{subsub:naive} an approach based on standard
  finite-size scaling, and
underline its limitations. In Sec. \ref{subsub:logs}, we discuss how
this approach can be generalized at the critical dimension, present
numerical results, and discuss a recent attempt to compute $P(n;A)$
exploiting duality. Although we focus ond
  comparing the scaling theory with results from the voter model with
  speciation, we remark that the theoretical approach presented
  in this section is more general and can be applied to a vast class
  of models at the critical dimension. 

\subsubsection{Power-law scaling relation}\label{subsub:naive}

In the voter model with speciation, the SAD is not only a function of the system size
$A$, but also of the speciation rate $\nu$.  Although we are mainly
interested in $2D$, it is instructive to consider the general case in
which $A=L^D$, where $L$ is the linear size of the sample. Following
\cite{PRL-Zillio,azaele2016statistical}, we assume a standard scaling
form for the SAD
\begin{equation}
  P(n;A,\nu) = n^{-\beta} \Psi(n \nu^\alpha, A \nu^{D/2})\,
\label{power-law}
\end{equation}
where the exponents $\alpha$ and $\beta$ remain unspecified for the
time being, whereas the exponent $D/2$ stems from the diffusive nature
of neutral models $\nu \sim t^{-1} \sim L^{-2} \sim A^{-2/D}$.  Note
that in models with long-range, non-diffusive dispersal
\cite{Rosindell2009} the scaling form might
differ. Equation~(\ref{power-law}) describes a power-law dependence of
$P$ on $n$, holding up to a scale determined by the scaling function
$\Psi$, that depends on dimensionless combinations of the population
size $n$, the speciation rate $\nu$, and the system size $A$.  To the
best of our knowledge, there is no available analytical prediction for
the exponent $\beta$.  The exponent $\alpha$ can be estimated in the
dual formulation of the voter model with speciation, where the
population size $n$ is the number of coalescences that occur before an
annihilation (see Sec. \ref{sec:duality}). This implies that $\alpha$
is the same exponent characterizing the temporal decay of the
density of coalescing random walkers, $\rho(t) \sim t^{-\alpha}$.
However, $\rho(t)$ decays as $\rho(t) \sim t^{-\min(1,D/2)}$ for
$D\neq 2$ and $\rho(t) \sim \log(t)/t$ in $D=2$, see
  eq. (\ref{bram}) and \cite{MF2,MF}. Consequently, one should expect
the power-law scaling of Eq.~(\ref{power-law}) to hold in $D=1$ and
$D\geq 3$, but not at the critical dimension $D=2$, where logarithmic
corrections should appear.

\subsubsection{Generalized scaling relation}\label{subsub:logs}

In order to allow for logarithmic corrections, Zillio \textit{et al.}
\cite{PNAS-Zillio} proposed the generalized
scaling relation
\begin{equation}
 P(n;A) = g(A) \Psi(n / f(A))\ .
\label{pnas}
\end{equation}
The dependence on $\nu$ was omitted as the above scaling law was
applied to observational data for which the speciation rate is unknown
and assumed to be fixed. The key aspect of Eq.~(\ref{pnas}) is that
$f$ and $g$, are general functions and not necessarily power-laws as
in conventional scaling, allowing for the possibility to include
logarithms or other functional dependencies. The scaling function
$\Psi(x)$ is still assumed to be a power law
\begin{equation}
\Psi(x) \sim x^{-\Delta}
\label{Delta}
\end{equation}
for small values of $x$, where $\Delta$ is an exponent to be
determined. Thus, also Eq.~(\ref{pnas}) postulates a power-law
dependence on $n$, but with a more general cut-off for large
areas. After specifying the functions $f$ and $g$, Eq. \ref{pnas} can
be tested by plotting $P(n;A)/g(A)$ versus $x = n/f(A)$ for a set of
different areas and assessing the quality of the data collapse onto a
single curve, $\Psi(x)$.

To determine the functions $f$ and $g$, we impose that $ P(n;A)$
has to be normalized,
$\int_{n_0}^\infty dn ~ g(A)~\Psi(n / f(A)) = 1$, and that its average
value has to be
$\langle n \rangle=\int_{1}^\infty ~dn ~ n ~ g(A)~ \Psi(n / f(A)) $.
Substituting the scaling form (\ref{Delta}) into these two equations,
it is possible to derive conditions that the functions $f$ and $g$
must obey, depending on the value of $\Delta$. In particular, the case
$\Delta=1$ is marginal and needs to be treated with care (other values
$\Delta \neq 1$ are analyzed in the Appendix). Approaching such a
  limit as $\Delta =1 - \epsilon$ with $\epsilon\ll1$,
Eq.(\ref{Delta}) becomes
\begin{equation}
\Psi(x)= x^{-1+\epsilon} \sim
\frac{1}{x}\big[\exp(\epsilon) \ln(x)\big] \sim \frac{1}{x} [ 1 +
  \epsilon \ln(x)]
\label{psi}
\end{equation}
up to first order in $\epsilon$. At the same order in $\epsilon$, the
two conditions for $P(n;A)$ become
$1 \sim g(A) f(A) \ln( f(A)) [ 1 + \frac{\epsilon }{2} \ln( f(A))] $
and $\langle n \rangle \sim g(A) f(A) ^2$, respectively, from which we
finally obtain
\begin{eqnarray}
   f(A) &=& \langle n \rangle \ln \langle n \rangle \left[ 1 + \frac{\epsilon
    }{2} \ln \langle n \rangle\right] \nonumber\\
g(A) &=& \frac{1}{\langle n \rangle \ln^2 \langle n \rangle \left[ 1 + \frac{\epsilon
    }{2} \ln \langle n \rangle\right]^2 }
\label{logs}
\end{eqnarray}
up to first order in $\epsilon$. Notice that both functions $f$ and $g$ include logarithmic
corrections. By means of a similar calculation, one can estimate the
$k$-th moment $\langle n^k \rangle$, and verify that all the moment
ratios $\langle n^{k} \rangle /\langle n^{k-1} \rangle $ scale in the
same way, up to a multiplicative constant
\begin{equation}
\frac{\langle n^{k} \rangle }{\langle n^{k-1} \rangle } 
=\frac{\int dn ~n^k ~ P(n;A)}{\int dn ~n^{k-1} ~ P(n;A)} \propto f(A)\qquad
k\ge 1 \,.
\label{eq:ratios}
\end{equation}
revealing a highly anomalous scaling.

Zillio \emph{et al.}~\cite{PNAS-Zillio}, showed that this scaling form
provides a much better collapse of empirical data from the Barro
Colorado tropical forest than a power-law scaling relation such as
Eq.~(\ref{power-law}). This supports the idea that $\Delta$ is close
to its marginal value $1$ in tropical forests.

\begin{figure}
 \includegraphics[width=9cm]{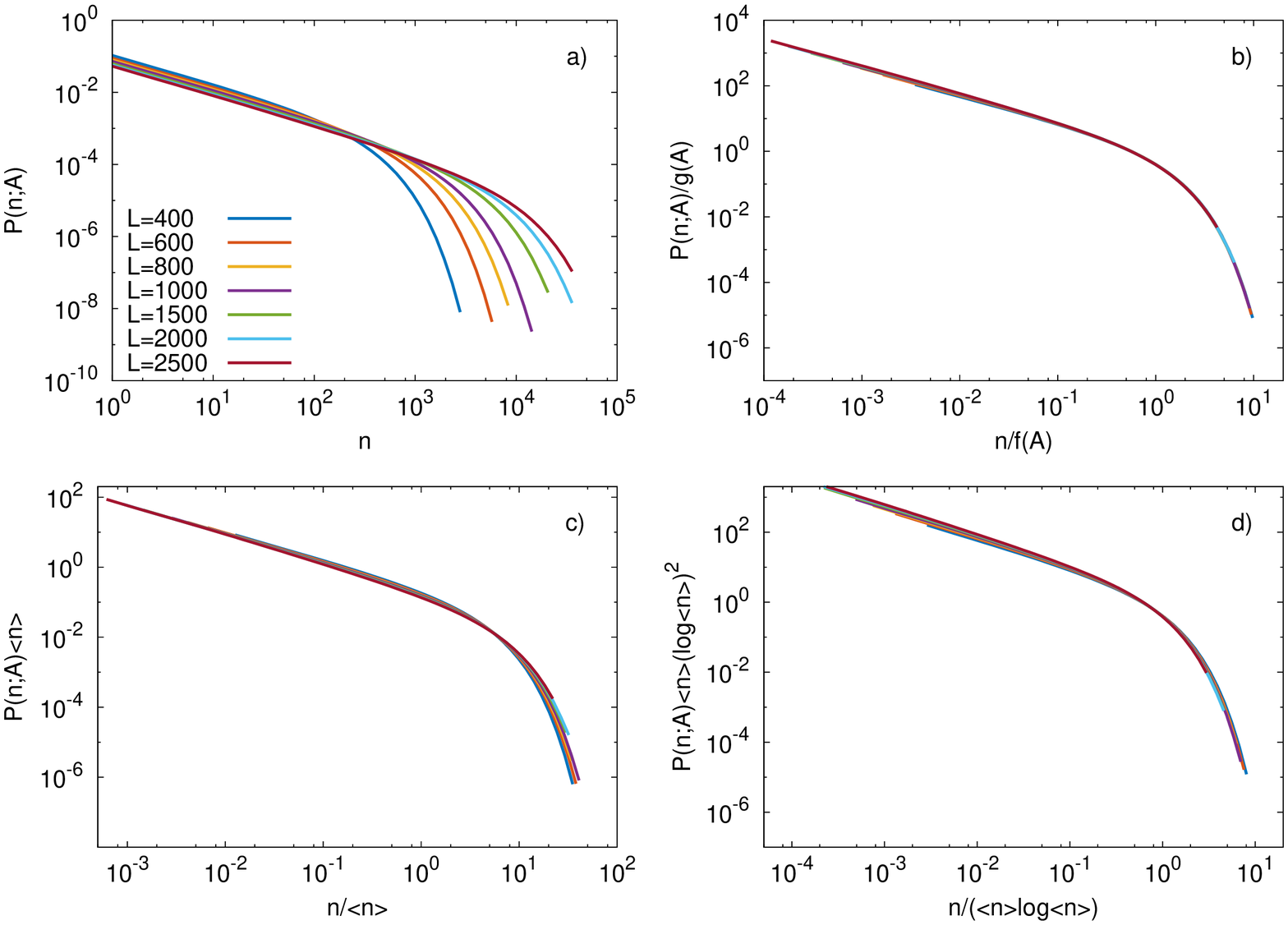}
 \caption{SAD and 
     data collapse. Results are presented for different linear
   system sizes and different speciation rates $\nu$, keeping the
   product $A \nu =200$ constant.  a) SADs for different linear sizes
   from $L=400$ to $L=2500$. b) Collapse of SADs by means of
   Eqs.(\ref{pnas}) and (\ref{logs}). The fitted parameter in the
   functions $f$ and $g$ is $\epsilon=0.08$. c) Naive collapse
   without logarithmic corrections, where deviation for perfect
   collapse are evident. d) Collapse with the scaling form of
   Eqs.(\ref{pnas}) and (\ref{logs}), but setting
   $\epsilon=0$. Also in this case the discrepancy is evident.}
\label{fig:panel}
\end{figure}

\begin{figure}
  \begin{center}
  \includegraphics[width=8cm]{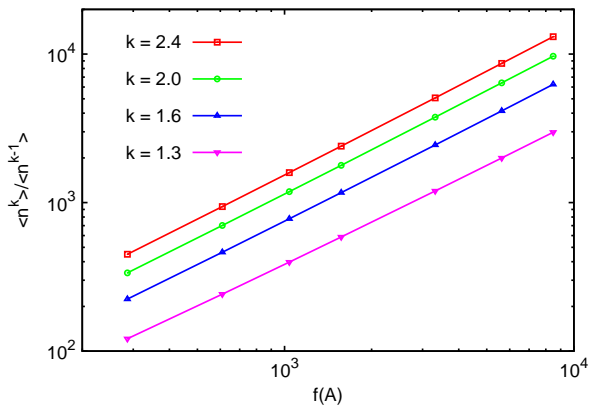}
  \caption{Moment ratios for different values of
    $k$. As predicted by Eq.(\ref{eq:ratios}), in the case $\Delta \approx 1$ all moment
    ratios $\langle n^k \rangle/ \langle n^{k-1} \rangle$ scale in the
    same way with $f(A)$ up to a multiplicative constant. As in
    Fig. \ref{fig:panel}, the fitted value is $\epsilon=0.08$. }
\label{fig:ratios}
\end{center}
\end{figure}

We tested computationally whether Eqs.  (\ref{pnas}) and (\ref{logs})
provide a good collapse of SADs obtained from the voter model with speciation
and whether the relationship between the moments,
Eq.(\ref{eq:ratios}), holds.  In simulations, an additional
parameter is the speciation rate $\nu$. As discussed above, $\nu$
appears in scaling relationships via the dimensionless combination
$A\nu^{D/2}$, that in $2D$ equals $A \nu$.  Thus, although
Eqs. (\ref{pnas}) and (\ref{logs}) do not include speciation
explicitly, we expect these relationships to hold if $A\nu$ is kept
constant. We therefore performed computational analyses fixing
$A \nu=200$, although the conclusions are robust against this choice.
Results are summarized in Figure \ref{fig:panel} which shows plots of
the SAD, for systems with different linear size, $L$ and different
speciation rates $\nu$ (with $L^2 \nu=A \nu=200$).  Observe in
Fig. \ref{fig:panel}a that the smaller the size (or the larger the
speciation rate) the smaller the maximal
abundance. Figure~\ref{fig:panel}b show the data collapse as given by
Eqs. (\ref{pnas}) and (\ref{logs}), where $\langle n \rangle $ is the
average number of individuals measured in each area $A$ and $\epsilon$
is a free parameter that we fitted obtaining $\epsilon= 0.08$ and
a remarkable collapse of the different
curves. The small value of $\epsilon$, is consistent with the assumed
small deviation from $\Delta=1$. A similar collapse for $A \nu=20$
leads to an even smaller value $\epsilon \approx 0.069$ (not shown).
We verified that either removing all logarithmic corrections (thus
plotting results as a function of $\langle n \rangle$) or simply
fixing $\epsilon=0$ in Eq. (\ref{pnas}) and (\ref{logs}) leads to less
convincing collapses, as shown in Fig. \ref{fig:panel}c and
\ref{fig:panel}d, respectively.  Clearly, these deviations can pass
unnoticed in the presence of statistical fluctuations. Probably, this
is the reason why in \cite{RC} a simple scaling law was claimed to
hold for the $2D$ voter model with
speciation. Finally, we also verified that moment ratios scale as
$f(A)$, as predicted by Eq.(\ref{eq:ratios}) and illustrated in
Fig.\ref{fig:ratios}.

\begin{figure}[htb!]
\begin{center}
\includegraphics[width=9cm]{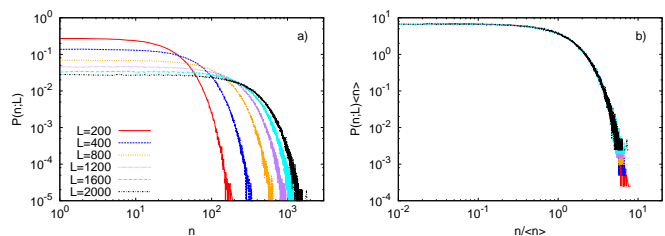} 
\caption{Species Abundance Distribution (SAD) in the $D=1$ voter model with
  speciation. a) SAD $P(n;L)$ vs $n$ at varying the system size $L$
  as labelled with $A\nu=40$ constant. b) Collapse of curved in (a)
  obtained withe rescaling SAD $P(n;L)\langle n\rangle$ vs $n/\langle
  n\rangle$. 
\label{fig:SAD1D}}
\end{center}
\end{figure}

In summary, a non-standard scaling form, including logarithmic
corrections, provides an excellent collapse both for empirical data
and for numerical simulations of the $2D$ voter
model. We remark that the scaling theory is
  phenomenological, and the small parameter $\epsilon$ controlling the
  importance of logarithmic corrections is, at this level, a
  non-universal free parameter. These results are in sharp contrast
  with the one-dimensional case, where logarithmic corrections are not
  expected. Indeed, Fig. (\ref{fig:SAD1D}) shows that the naive
  scaling form $P(n;A) \langle n \rangle$ vs. $n/\langle n \rangle$
  (derived in Appendix A for the case $\Delta \neq1$) yields a perfect
  collapse for SADs in one-dimensional systems.

  It is interesting to remark that the data collapsed in
  \cite{PNAS-Zillio} were obtained from tropical forests of different
  areas $A$. It is reasonable to assume that the speciation rate $\nu$
  do not vary much among these forests. Therefore, the product $A\nu$
  is not fixed, as in our computational analyses. A possible
  explanation is that, although the collapse achieved in this way is
  not perfect, the deviations from perfect scaling are too small to be
  appreciated in observational data due to the limited sample size. We
  have verified in simulations (not shown) that
  keeping $\nu$ constant (rather than $A \nu$ constant) small
  deviations from perfect collapse are observed.

We conclude this section mentioning that a heuristic expression for
the SAD has been recently derived for the voter model with speciation
following a completely different approach
\cite{danino2016spatial,Danino2}. Let us define $P(x,t)$ as the
distribution of the number of individual of a given species at time
$t$. If we approximate $x$ as a continuous quantity, we can
heuristically write a Fokker-Planck equation for the evolution of
$P(x,t)$
\begin{equation}\label{shnerbfp}
\partial_t P(x,t)=\nu\partial_x [xP(x,t)]+\partial^2_x[I(x)P(x,t)]
\end{equation}
where the first term in the right hand side is the negative drift due
to speciation, and the second is the fluctuation in population size,
where $I(x)$ is the average number of interfaces of a species of size
$x$. The crucial underlying approximation is to neglect fluctuations
of $I(x)$, which is appropriate if the distribution
of the number of interfaces at fixed value of $x$ is a very peaked
function. In this simple framework, all the dependence on the spatial
dimension of the voter model is recap into the function $I(x)$. The
steady-state solution of Eq.~(\ref{shnerbfp}) is
\begin{equation}\label{shnerbss}
  P_{st}(x)=\frac{e^{-\nu\int dx \frac{x}{I(x)}}}{I(x)} .
\end{equation}
From duality considerations \cite{danino2016spatial,Danino2}, the average
number of interfaces must scale in $2D$ as $I(x)=x/(1+c\ln x)$ where
$c$ is a non-universal constant. Notice how the expression of $I(x)$
includes familiar logarithmic terms and that the
  constant $c$ plays the role of the exponent $\epsilon$ in the
  scaling theory. Substituting this expression into
Eq.~(\ref{shnerbss}) leads to an explicit expression for the SAD,
which obeys a scaling law with logarithmic corrections similar to
Eq.~(\ref{logs}), though not identical. A more detailed comparison between
this result and the previous scaling form is an interesting issue, but
beyond the scope of this review.

\subsection{Species persistence-times}

So far, we have considered neutral predictions of
  static ecological observables. However, neutral theory can also be
  used to predict time-dependent properties.  A chief example is the
  distribution of survival times. The survival time $\tau$ (also
  called "persistence time") within a geographic region is defined as
  the time occurring between the speciation event originating a given
  species and its local extinction
  \cite{pigolotti2005species}. Recent empirical work on north-american
  birds and herbaceous plants revealed that the probability of
  observing a persistence time $\tau$ decays as as power laws
  $P(\tau) \sim \tau^{-1.83}$ and $P(\tau) \sim \tau^{-1.78}$
  respectively, with area-dependent exponential cut-offs
  \cite{Bertuzzo1,Bertuzzo2}.

In the voter model with speciation, the survival
  probability as a function of time can be computed analytically.
  Also in this case, the calculation relies on duality 
  \cite{bramson1991,MF,Lee}. In $2D$ and in the limit of vanishing
  $\nu$ one obtains
\begin{equation}
P(\tau) \sim \frac{\ln \tau}{\tau^2} \, 
\label{log}
\end{equation}
while standard power-law scaling $ P(\tau) \sim \tau^{-1/2}$ is
expected in 1D. For non-negligible values of $\nu$, these scaling
forms are cut-off by a $\nu$-dependent exponential factor
$\exp (-\nu \tau)$ in either dimension. Also in this case, diffusive
scaling relates the characteristic time scale $1/\nu$ with a length
scale $\xi$ via $\xi\sim\sqrt{\nu}$. This explains the aforementioned
area-dependent cut-offs observed in empirical data \cite{Bertuzzo1}.

\begin{figure}[htb]
\begin{center}
\includegraphics[width=9cm]{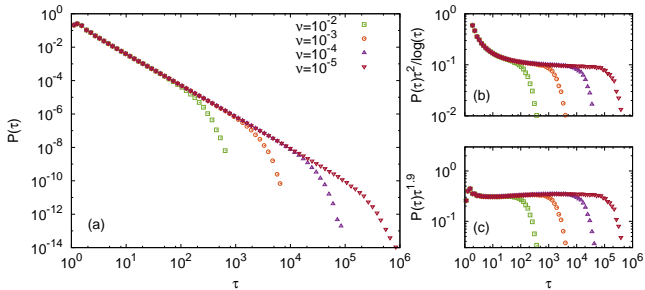}
\caption{Species persistence times. (a) Probability distribution
  function of species persistence times for different values of the
  speciation rate $\nu$ as in label.  (b) and (c) show the pdf
  rescaled with the logarithmic correction, $P(\tau) \tau^2/\ln\tau$,
  and with a power law, $P(\tau) \tau^{1.9}$,
  respectively. 
  \label{fig:persistence}}
\end{center}
\end{figure}

Species persistence times in simulations of the $2D$ voter model with
speciation are shown in Fig. \ref{fig:persistence}a. Panels (b) and
(c) show compensated plots of the simulation results. The simulations
support the prediction of eq. (\ref{log}) (panel b), and also
illustrate that a power law with an exponent close to $2$ ($1.9$ in
this case) provides a good approximation of the scaling predicted by
Eq. (\ref{log}) in a broad range of scales (panel c), consistently
with the empirical findings in \cite{Bertuzzo1,Bertuzzo2}.

\section{Other neutral models\label{sec:otherneutral}}

In the voter model with speciation, the habitat is saturated and each
site is always occupied by an individual.  In this section, we study
neutral spatial models where the number of individuals that can
inhabit a site is varied.  We consider three
variants: the stepping-stone model with speciation,
where each site can host many individuals but the landscape
remains saturated; the contact process with speciation, where
occupancy is limited to a maximum of one individual per site, but
sites can also be empty; and the O'Dwyer-Green model, where occupancy
is unbounded.

\subsection{Stepping-Stone Model with speciation\label{sec:ssm}}

In the voter model, each lattice site hosts a single
individual. This assumption is appropriate for big sessile
species, such as trees, where each individual occupies a
  well-defined area and exploits its local resources.  On the other
side of the spectrum, microorganisms, such as small eukaryotes or
bacteria, are often present in very large numbers on tiny spatial
scales, where all individuals share the same resources. For these
species, it is more appropriate to think of the habitat as subdivided
into small patches, connected by migration and each hosting a large
number of individuals directly competing with each other
\cite{Fenchel2004-2}.  To model such ecological cases, in
this section we consider the stepping-stone model
\cite{kimura1964stepping,korolev2010genetic} with speciation, which
generalizes the voter model with speciation to the case in which each
site hosts a fixed number $M$ of individuals.

Similar to the voter model with speciation, at each time step an
individual is randomly chosen and killed. With probability $\nu$, it
is replaced by an individual of a novel species. With complementary
probability $(1-\nu)$, a reproduction event occurs. The parent of the
new individual is selected with probability $(1-\mu)$ among the
surviving $M-1$ individuals present at the same site, and with
probability $\mu$ among the $M$ individuals in a randomly chosen
neighboring patch (according to a probability distribution on the
neighbors $P(\vec{r})$, similar to the case of the voter model). The
particular case of $M=1$ reduces to the voter model with speciation up
to a time rescaling $t\rightarrow \mu t$.  Like the voter model, the stepping-stone
model admits a dual representation in terms of coalescing random
walkers with annihilation, which can be exploited for efficient
numerical simulations. The main difference with respect to the
  dual of the voter model is that, in the dual stepping-stone model,
at each step a random walker can either move to another site or stay
in the site of origin. Coalescence can happen in both circumstances,
corresponding to reproduction of an individual from neighboring sites
or from the same site. For full details on the implementation we refer
to \cite{cencini2012ecological}.

As revealed by numerical simulations of the stepping-stone model based
on the dual representation, SARs are qualitatively similar to those of
the voter model, although the exponents $z$ are, in general, smaller
than in the voter model \cite{cencini2012ecological}. In particular,
the exponent depends not only on $\nu$, but also on the combination of
parameters $M\mu$, which determines the regimes of the model. For
$M\mu\ll 1$, each local site is likely to contain only one species. In
this limit, each site behaves as one individual up to a time
rescaling, so that one should expect the same exponents as in
the voter model with speciation.  In the opposite limit $M\mu\gg 1$,
there is a large diversity of species at each site. An analytical
argument suggests that, in this latter limit, the exponent should be a
factor two smaller than in the former limit
\cite{cencini2012ecological}. Let us study the limit
  $M\mu\gg 1$ in the dual representation. Since random walkers in the
same site have a low probability of coalescence, they will wander for
a long time before coalescing. Therefore, we can assume that,
asymptotically, they will behave as in the well-mixed case.  This
implies that their density in an area smaller or equal than $\xi^2$
approximately decays according to the mean-field formula
\begin{equation}
\rho(t)\sim \frac{1}{t}\ .
\end{equation}
Observe that in this case the characteristic length is
$\xi=\sqrt{\mu/\nu}$, as random walks diffuse with
probability $\mu$ at each time step.  Proceeding as in Eq.~ (\ref{eq:sxi_vm}), the
average number of species in an area $\xi^2$ can be estimated as
\begin{equation}\label{eq:sxi_ssm}
S(\xi^2)\sim \nu M\xi^2\int_{t_0}^{\mu/\nu=\xi^2} \ \frac{dt}{t} =
M\mu\ln\left[\frac{\xi^2}{t_0}\right] .
\end{equation}

\begin{figure}[htb]
\begin{center}
  \includegraphics[width=8cm]{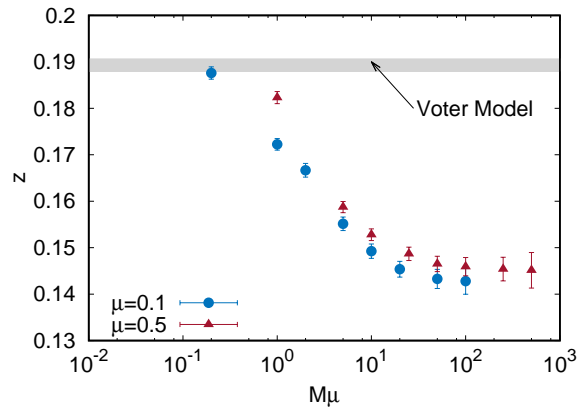}
  \caption{Species-area exponents for the Stepping Stone Model at
    fixed $\nu=10^{-6}$, different local population size $M$ and
    dispersal rate $\mu$, with NN dispersal. The
    numerical estimate of the exponent $z$ in the voter model for
    NN-dispersal and the same value of the speciation rate is shown
    for comparison. \label{fig:ssm}}
\end{center}
\end{figure}

To compute $z$, we also need an estimate for $S(1)$, that in this case
is not trivially equal to one. As the population is assumed to be
well-mixed in an area equal to $\xi^2$ or smaller, the composition of
a single site can be thought as a sample of $M$ individuals from this
well-mixed population. The probability distribution of the abundance
in such a sample is given by Ewens' sampling formula
\cite{Ewens}. Substituting its expression yields
\begin{equation}\label{eq:s1_ssm}
S(1)=\sum\limits_{j=0}^{M-1} \frac{M\mu}{M\mu+j}\approx M \mu \ln
(1+\mu^{-1})\ .
\end{equation}
Combining Eqs. (\ref{eq:sxi_ssm}) and (\ref{eq:s1_ssm}) and assuming a
power law in the range from $A=1$ to $A=\xi^2$, we find an exponent
\begin{equation}
z\sim \frac{\ln(\xi^2)}{\ln \ln(\xi^2)}=
\frac{\ln\ln(\nu/\mu)}{\ln(\nu/\mu)}
\label{estimate:ssm}
\end{equation}
which, to the leading order, is a factor $2$ smaller than the
corresponding estimate for the voter model (\ref{eq:dl}). The decrease
of the exponent $z$ with the combination of parameters $M\mu$ is
confirmed in numerical simulation, see Fig. \ref{fig:ssm}, although
the asymptotic reduction is less than the factor two predicted by the
approximate estimate of eq. (\ref{estimate:ssm}).

Summarizing, the stepping-stone model at large local community size
$M$ yields smaller values of the species-area exponent $z$ than the
voter model \cite{cencini2012ecological}. This fact is consistent with
the ecological observation that microbial communities, characterized
by very large local community sizes, typically display very shallow
species-area relations, and that in general there seems to be a
positive correlation between the exponent $z$ and the body size of a
taxonomic group \cite{horner2004}. In the stepping-stone model, a
decrease in the SAR exponent is observed in the regime
$M\mu\gg 1$ where each site hosts a large number of species and
therefore provides a buffer for biodiversity
\cite{cencini2012ecological}. This interpretation is also consistent
with the ``cosmopolitan'' nature of many microbial species, i.e. the
fact that relatively small communities of microbes host a biological
diversity comparable with that observed in the whole planet
\cite{Fenchel2004-1,Fenchel2004-2}. This feature has sometimes been
explained invoking the fact that microbes have the possibility of
long-range dispersal \cite{Fenchel2004-1}. However, numerical
simulations show that, in the voter-model with speciation, long-range
dispersal leads to steeper, rather than shallower SARs \cite{Rosindell2009}.

\subsection{Contact Process with speciation}

In the voter model, every dead individual is instantly replaced by a
newborn, leading to a constantly saturated environment. The implicit
underlying assumption is that the birth rate is infinite, so that
death events are the rate-limiting steps. Such assumption
constitutes a good approximation in resource-rich ecosystems. In
less rich ecosystems, where the birth rate is finite, the environment
is not always saturated and empty gaps can exist
\cite{loreau2000}.

To explore this latter case, we study here the contact process with
speciation, which is the multi-species
variant of the well-known contact process
\cite{griffeath1981,Liggett1975,durrett1994stochastic,MarroBook,Hinrichsen2000}. As
usual, we consider the model on a $2D$ square lattice. Sites of the
lattice can be occupied by individuals belonging to different species
or empty. The model is defined in continuous time; each individual
dies at a rate $d$ and reproduces at a rate $b$. In case of a death,
the site is simply left vacant. A reproduction event is considered
successful only if the individual has at least one vacant neighboring
site. In such a case, one of the vacant neighboring sites is
chosen at random. With probability $\nu$, the site is occupied by an
individual of a new species (speciation event); with complementary
probability, $(1-\nu)$, the newborn is of the same species as the
parent.

As in the standard contact process
\cite{griffeath1981,Liggett1975}, the parameter determining
the asymptotic density of occupied sites $\rho$ is the dimensionless
birth-to-death ratio $\eta=b/d$. For
$\eta<\eta_c\approx 1.649$ the absorbing state in which all sites are
empty is stable. A non-equilibrium phase transition at $\eta=\eta_c$
separates this region from a stable active phase ($\eta>\eta_c$)
characterized by a non-vanishing value of $\rho$ that depends on
$\eta$ \cite{MarroBook,Hinrichsen2000}. For $\eta\rightarrow\infty$
one has $\rho\rightarrow 1$ and the model is equivalent to the voter
model with speciation \cite{durrett1994stochastic}.

\begin{figure}[htb]
\begin{center}
\includegraphics[width=8cm]{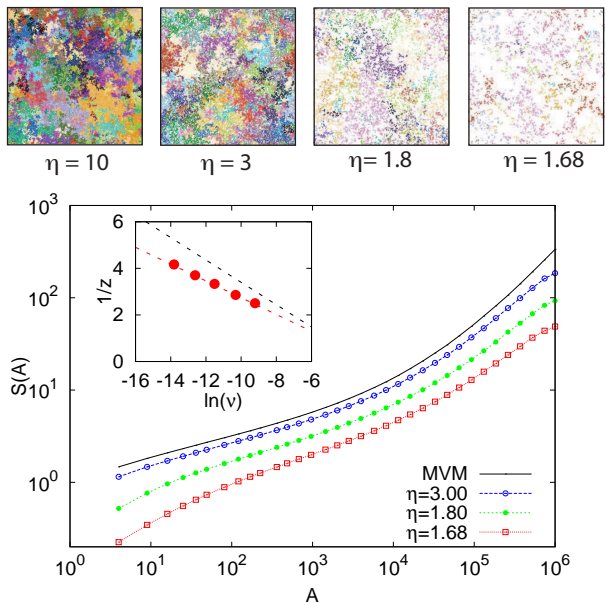}
\caption{(top) Snapshots of configurations of the contact process with
  speciation at different values of the birth-to-death rate ratio
  $\eta$ and $\nu=10^{-4}$. In each panel, each color represents a
  different species. (bottom) SARs at different
  values of the birth-to-death rate $\eta$ (shown in the figure
  legend) and $\nu=10^{-5}$. (inset) Red dots: estimated exponent $z$
  as a function of $\nu$ for the contact process with speciation at
  $\eta=1.68$. Red dashed line is a linear fit; black dashed line is
  the corresponding fit for the voter model with speciation for
  comparison. We have chosen a NN dispersal kernel in
  all panels. \label{fig:cp}}
\end{center}
\end{figure}

The CP is  a self-dual model. Therefore,  duality cannot be
  exploited in numerical simulations as in the case of the voter model.  Forward
simulations show that the SAR and the corresponding exponents are
remarkably similar to the voter model with speciation even at small
values of $\eta$, corresponding to very fragmented ecosystems as shown
in Fig. \ref{fig:cp}. For values of $\eta$ very close to $\eta_c$ (but
within the active phase) and small values of $\nu$, SAR exponents tend
to be smaller than in the voter model, see inset of Fig. \ref{fig:cp}.

In principle, in a very fragmented ecosystem it would not make sense
to sample empty areas, or areas with very few individuals. With this
idea in mind, an alternative to the standard definition of SAR used so
far is to weigh the sample of a given area with its
number of individuals, i.e. of occupied sites. Adopting this
definition one finds qualitatively different SARs for small values of
$\eta$ \cite{cencini2012ecological}.  In particular, these SARs do not
seem to be characterized by a clear power-law range. We refer the
reader to Ref. \cite{cencini2012ecological} for a broader discussion
of this issue.

\subsection{O'Dwyer-Green model}

We have seen that finding exact results for neutral spatial models
constitutes a formidable problem, and even in the simple case of the
voter model only asymptotic results are known.

To make progress in this direction, O'Dwyer and Green proposed a
spatial neutral model in which individuals do not compete, i.e. the
site occupancy is not bounded \cite{o2010field}. In their model, each
individual can reproduce at a rate $b$, giving rise to a newborn
located according to a dispersal distribution, die at a rate $d$, or
speciate at a rate $\nu$, giving rise to a newborn of a new
species. The model was studied at the critical point $b+\nu=d$. The
lack of interaction considerably simplifies the mathematical
treatment: the model can be mapped into a field theory from which the
authors of \cite{o2010field} obtained an analytical expression for the
species-area law and the dependence of $z$ on $\nu$. In particular,
the solution was derived by writing an equation for the distribution
of a generic species, which was solved by imposing detailed balance.
However, Grilli and coworkers \cite{grilli2012absence} pointed out a
flaw in this procedure. In this model all species are transient, as
the birth rate of each species is always smaller
than the death rate because of speciation. This implies
that all species eventually go extinct, so that the detailed balance
(i.e. equilibrium) assumption is not valid.

An often overlooked aspect of the O'Dwyer and Green model is the lack
of a carrying capacity. Although well-mixed neutral
  models commonly do not have a carrying capacity (beside that of the
  entire ecosystem), a local carrying
  capacity, i.e. a maximum occupancy of each lattice site, is
  a standard ingredient in spatial neutral theory, shared by all models we discussed so far.
 In the O'Dwyer and Green model, since the dynamics of the entire ecosystem is
a critical branching process, the population at each site undergoes
huge fluctuations.  This fact implies as a drawback that numerically
simulating the steady-state of the model and sampling its
configurations is extremely difficult. While the authors of
\cite{grilli2012absence} clearly pointed out that the detailed balance
solution leads to several inconsistencies and is therefore not valid,
to the best of our knowledge there have been no attempt of comparing
this solution with numerical simulations to see if detailed balance
can provide a reasonable approximation of the dynamics in some
particular regimes or limits.

Currently, the research of spatial neutral models that can be solved
analytically is still open \cite{o2017cross}. In this
direction, although this review focuses on lattice models, we mention
a recent phenomenological attempt based on a spatial Fokker-Planck
equation where both space and population sizes are continuous
variables \cite{azaele2016phenomenological}.

\section{Near-neutral models \label{sec:nearneutral}}

In the previous sections, we focused on neutral
  ecological models. However, in real ecosystems the neutral
  assumption is (at best) a crude approximation. It is thus
  interesting to examine some of the main effects of non-neutral
  forces, also because many biodiversity patterns
  that are well predicted by neutral models are also found in richer,
  non-neutral models \cite{McGill2003,tilman2004,GilbertLechowicz}.
  A main difficulty in comparing neutral and non-neutral models is the
  large number of possible ecological effects (and corresponding
  parameters) that typically enter the latter.  In
    this section, with the aim of understanding basic non-neutral
    effects in a simple setting, we present a minimal model introduced
    in \cite{PC2010}, where one can continuously move from a neutral
    to a non-neutral scenario by varying a single parameter, tuning
    the amount of spatial disorder. We then discuss generalizations to
    other types of spatio-temporal disorder.

\subsection{Habitat-preference model}

We consider a variant of the voter model where
different sites are preferred habitats for each one of the competing
species.  For the sake of simplicity, we limit ourselves to the case
of two species $A$ and $B$ with $N_A$ and $N_B$ individuals,
respectively. We assume habitat saturation, so that the total
population is $N=N_A+N_B=L^2$ where the system is a square lattice of
size $L$ with periodic boundary conditions. Individuals of type $A$
and $B$ can also migrate to the system from an infinite reservoir
where they are equally
  represented. Each lattice site can be of type $a$ or $b$,
i.e. being a preferred habitat for colonization by species $A$ or $B$,
respectively.  After colonization, mortality and dispersal do not
depend on being on a preference site. Ecologically, this means that
the fitness advantage belongs to the seeds and not to the individuals
themselves (see \cite{Chesson1981} for a different choice).  The $a$
vs $b$ character of each site is chosen randomly at the beginning and
it remains fixed over time -- quenched disorder. To maintain the model
globally symmetric, we assume equal proportions of $a$ and $b$ sites
and that intensity of the two biases ($a$ favoring $A$ and $b$
favoring $B$) are identical. The dynamics proceeds as follows. At
each discrete time step, a lattice site is randomly chosen with
uniform probability and the residing individual is killed.  The
individual is replaced either by an immigrant from the reservoir (with
probability $\mu$) or by an offspring of an individual residing in one
of the four neighboring sites (with probability $(1-\mu)$). In both
cases, the colonization probability is biased by an additional factor
$\gamma$ for the individuals that have preference for the empty site.
In formulas, the probability of colonization of a site $x=\{a,b\}$ by
an individual $X=\{A,B\}$ ($Y=\{B,A\}$) having (not having) preference
for that site is
\begin{equation}
\label{eq:ratesNN}
\begin{array}{ll}
  W^x_X(n_X,n_Y)=&(1-\mu)\frac{(1+\gamma)n_X}
{(1+\gamma)n_X+n_Y}+\mu\frac{1+\gamma}{2+\gamma}\\
&\\
  W^x_Y(n_X,n_Y)=&(1-\mu)\frac{n_Y}{(1+\gamma)n_X+n_Y}+
\mu\frac{1}{2+\gamma}\,,
\end{array}
\end{equation}
respectively, where $n_X$ ($n_Y$) denotes the number of individuals of
species $X=\{A,B\}$ ($Y=\{B,A\}$) in the neighborhood of the
considered site. Similar models have been proposed also in the context
of heterogeneous catalysis \cite{Redner1} and social dynamics
\cite{Redner2}.  For $\gamma=0$ and $\mu=0$, the standard (neutral)
voter model with two species is recovered. For $\gamma=0$ but
$\mu\neq 0$, it corresponds to the noisy voter model
\cite{Kirman93,GranovskyMadras95}.

Also in this model, the results can depend on the choice
  of the dispersal kernel $P({\bf r})$. Here we focus on the NN
  dispersal and global dispersal (GD), i.e. a mean-field version of
  (\ref{eq:ratesNN}). The GD case can be thought as a variant of the
two islands model \cite{Moran1962} of population genetics, where each
island host $N/2$ individuals and is favorable to one of the two
species. In the mean-field version, the state of the system is
univocally determined by the numbers of individuals $N_{Aa}$ and
$N_{Bb}$ residing on their island of preference. The numbers of
individuals outside their island of preference are
$N_{Ba}=N/2-N_{Aa}$ and $N_{Ab}=N/2-N_{Bb}$. The dynamics is then
  fully specified by the probabilities per elementary steps that
$N_{Xx}$ (with $X=\{A,B\}$ and $x=\{a,b\}$) increases or decreases by
a unit:
\begin{eqnarray}
\label{eq:ratesGD}
  \mathcal{W}_{N_{Xx}\rightarrow N_{Xx}+1}&=& \left(\frac{1}{2}-\frac{N_{Xx}}{N}
                                              \right) \ W_X^x(N_A,N_B) \nonumber\\
  \mathcal{W}_{N_{Xx}\rightarrow N_{Xx}-1}&=& \frac{N_{Xx}}{N}\ 
                                              W_Y^x(N_A,N_B) 
\end{eqnarray}
where $W^x_Y$ and $W^x_X$ are given by eqs. (\ref{eq:ratesNN}) with
$n_X$ and $n_Y$ replaced by $N_X=N_{Xx}+N_{Xy}$ and
$N_Y=N_{Yy}+N_{Yx}$, respectively.

%%%%%%%%%%%%%%%%%%%%%%%%%%%%%%%%%%%%%%%%%%%%%%%%%%%%%%%%%%
\subsection{Extinction times}
%%%%%%%%%%%%%%%%%%%%%%%%%%%%%%%%%%%%%%%%%%%%%%%%%%%%%%%%%%

In the absence of immigration ($\mu=0$) and for finite populations
$N<\infty$, persistent coexistence of the two species is not possible:
demographic stochasticity eventually drives one of the species to
extinction (the absorbing state) with the \textit{fixation} (in the
jargon of population genetics) of the other species. In this
  case, information on the system can be obtained by studying the
dynamics toward extinction \cite{Chesson1981}. Of particular interest
is the average extinction time, $\langle T_{ext}\rangle$, and its
dependence on system properties, such as the deviation from neutrality
and the population size.

In the neutral case ($\gamma=0$), as discussed, the system recovers
the voter model with NN dispersal and the Moran model \cite{Moran58}
in the version with global dispersal.  In this limit, the extinction
time is set by the population size.  In particular, for large $N$ we
have $\langle T_{ext}\rangle \sim N\ln N$ for NN-dispersal
\cite{krapivsky1992} and $\langle T_{ext}\rangle\sim N$ for global
dispersal \cite{Moran58,Gillespie}. To inquire the effect of habitat
preferences we performed simulations of the model (\ref{eq:ratesNN})
with an initial condition $N_A=N_B=N/2$ until the
extinction of one of the two species.
\begin{figure}[htb]
\begin{center}
\includegraphics[width=8cm]{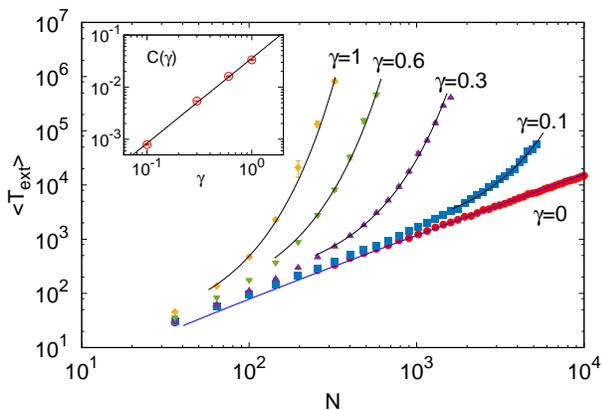}
\caption{Extinction times for the model with NN
  dispersal without immigration ($\nu=0$). Mean extinction time
  $\langle T_{ext}\rangle$ as a function of $N$ for different values
  of $\gamma$ as in label. The blue curve approximating the neutral
  $\gamma=0$ data points corresponds to the neutral expectation
  $\langle T_{ext}\rangle\propto N\ln N$, the black curves over the
  symbols for $\gamma\neq 0$ correspond to exponential fits of the
  form $\langle T_{ext}\rangle\propto \exp(C(\gamma)N)$. The inset
  shows (symbols) $C(\gamma)$ vs $\gamma$, while the black solid line
  display the best fit $C(\gamma)=A \gamma^\beta$ with
  $\beta\approx 1.63$.  The average extinction time is
  obtained by an annealed average, i.e. by randomizing the preference
  sites at each realization. Each point represents an average over $10^3$
  realizations.\label{fig:nearneutral_times}}
\end{center}
\end{figure}

Figure \ref{fig:nearneutral_times} shows the average extinction time,
measured in generations, i.e. $N$ elementary steps of
eqs. (\ref{eq:ratesNN}), as a function of the population size $N$ for
different values of $\gamma$. For
$\gamma=0$ we observe the $N\ln N$ behavior expected in the neutral
case. Habitat preference ($\gamma>0$) leads to a dramatic increase of
the average extinction time, which becomes exponential in $N$
\begin{equation}
\langle T_{ext}\rangle \propto \exp(C(\gamma) N)\,,
\end{equation}
for large enough $N$. The dependence of the constant $C(\gamma)$
on $\gamma$, shown in the inset, is well-fitted by a power-law with
exponent $\approx 1.63$.  The mean-field version of the model presents
similar qualitative features with the only difference that
$\langle T_{ext}\rangle \propto N$ for $\gamma=0$ and with some
differences in the $\gamma$ dependence of $C(\gamma)$, as shown in
\cite{PC2010}.

The exponential dependence of the average extinction times on $N$
indicates that habitat preference has a stabilizing impact on the population
dynamics. Indeed, when $N$ is large enough, the two species coexist on
any realistic time scale.  The stabilizing effect of habitat
preference reflects also in the probability of fixation $P_{fix}$,
i.e. the probability that a species, say $A$, gets fixated when
initially present as a fraction $x=N_A/N$ of the population. In the
neutral case, standard computation \cite{Gillespie} shows that
$P_{fix}(x)=x$.  As shown in \cite{PC2010}, when $\gamma$ is
increased, $P_{fix}(x)$ develops a much steeper dependence on $x$ and
quickly reaches values $\approx 1/2$ even for small $x$, provided
that $\gamma$ is large enough. In other words, the stabilization
due to habitat preference tends to compensate any initial
disproportion between the population of the two species.

%%%%%%%%%%%%%%%%%%%%%%%%%%%%%%%%%%%%%%%%%%%%%%%%%%%%%%%%%%
\subsection{Coexistence}
\begin{figure}[htb]
\begin{center}
\includegraphics[width=8cm]{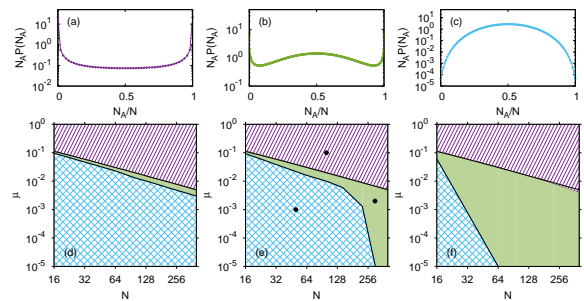}
\caption{Different regimes of coexistence for the case with
  NN dispersal and immigration for the model with
  habitat preference. Top panels show the stationary distribution
  $P(N_A)$ for $\gamma=0.3$ and (a) $N=50$ with $\mu=10^{-3}$, (b)
  $N=300$ with $\mu=2\times 10^{-3}$, and (c) $N=100$ with
  $\mu= 10^{-3}$, corresponding to a typical distribution in the cases 
  of monodominance, mixed regime and pure coexistence, see
  text. Bottom panels show how the three regimes partition the
  $N,\mu$-parameter space for different values of $\gamma$: (d)
  $\gamma=0$ corresponding to the neutral case, (e) $\gamma=0.3$ and
  (f) $\gamma=1$.  The three points in (e) correspond to the
  distributions displayed in the top panels, as labelled by the color
  coding. \label{fig:nearneutral_coexistence}}
\end{center}
\end{figure}

%%%%%%%%%%%%%%%%%%%%%%%%%%%%%%%%%%%%%%%%%%%%%%%%%%%%%%%%%%
In the presence of immigration ($\mu>0$), a locally extinct
species can recolonize, leading to a dynamical coexistence between the
two species.  However, if the typical recolonization time $1/\mu$ is
large compared to the average extinction time $\langle
T_{ext}\rangle$, such recovery from extinction is slow and
unlikely. Therefore, most of the time the ecosystem is
  dominated by one of the two species.
Therefore, the
distribution of the population size of any of the two species, $P(X)$
($X=A,B$) is peaked at $0$ and at the population size $N$,
corresponding to dominance of either of the two species. We denote
this regime as {\em monodominance}, see
Fig. \ref{fig:nearneutral_coexistence}a. In the opposite limit
$\langle T_{ext}\rangle\gg 1/\mu$, temporary extinctions are very
unlikely and the distribution is peaked at $N_A=N_B=N/2$ leading to
\textit{pure coexistence} of the two species
(Fig.~\ref{fig:nearneutral_coexistence}c). For intermediate values of
$\mu$, temporary extinctions are still possible though the
replenishment due to immigration will tend to equilibrate the two
populations. In this case of \textit{mixed coexistence}, the
distribution is characterized by three local maxima at $N_X=0,N/2,N$
(Fig.~\ref{fig:nearneutral_coexistence}b).

Figs.~\ref{fig:nearneutral_coexistence}d,e,f show the three
regimes of coexistence  in the $N-\mu$ parameter space for
the model with NN-dispersal for different habitat preference
strength $\gamma$ (increasing from left to right). In the mean-field
model, we find the same qualitative features, except that for
$\gamma=0$ the mixed regime is absent, so that one has a direct
transition from monodominance to pure coexistence \cite{PC2010}.

The main emerging feature is that increasing habitat preference
expands the region of parameter space corresponding to mixed
coexistence at the expenses of monodominance. Surprisingly, the pure
coexistence regime seems to be insensitive to the degree of habitat
preference. In particular, the critical line $\mu_c(N)$ separating it
from the mixed regime seems to be the same that separates
coexistence from monodominance in the neutral model ($\gamma=0$) with
global dispersal, which is given by the expression
$\mu_c(N)= 2/(2+N)$.  This result can be obtained in the
  following way. For $\gamma=0$, the transition rates
(\ref{eq:ratesGD}) can be expressed in terms of the rates for $N_A$ to
increase/decrease by one
\begin{eqnarray}
\label{neutralrates}
\mathcal{W}_{N_A\to N_A \pm 1}&=&
\frac{\frac{N}{2} \pm \left(\frac{N}{2}-N_A\right)}{N}\times\\
&\times&
\left[(1-\mu)\frac{\frac{N}{2} \mp \left(\frac{N}{2}-N_A\right)}{N}+
  \frac{\mu}{2}\right]\,.\nonumber
\end{eqnarray}
Then, the equilibrium distribution $P(N_A)$ can be computed imposing
the detailed-balance condition
\begin{equation}
\label{eq:db}
  \frac{P(N_A+1)}{P(N_A)}=\frac{\mathcal{W}_{N_A\to N_A+1}}{\mathcal{W}_{N_A\to N_A-1}}\,,
\end{equation}
which must hold at stationarity since the process is one dimensional
\cite{gardiner2009}.  To determine $\mu_c(N)$ for the transition from
monodominance to coexistence, it is sufficient to determine
whether, for small $N_A$, $P(N_A)$ is an increasing or a
  decreasing function.  Using (\ref{eq:db}) with (\ref{neutralrates})
and imposing $P(N_A+1)>P(N_A)$ one obtains, after some algebra, the
inequality $[(2+N)\mu-2](N-2N_A-1)>0$, which is verified whenever
$\mu>2/(2+N)$. Notice that, in the case of global dispersal, the
distribution is uniform along this line, i.e. for $\mu=\mu_c$ one
finds $P(N_A)=1/N$.

\subsection{Generalizations of the habitat-preference model}

To gain physical insight into the different regimes shown in
Fig. \ref{fig:nearneutral_coexistence}, a variant of the habitat
preference model was introduced and analyzed for the global dispersal
case in \cite{Borile-JSTAT1}.  By considering the first two terms of a
system-size expansion of the master equation, results in the
infinite-size limit and finite-size corrections were derived.  In the
infinite-size limit, i.e.  neglecting the effect of fluctuations, the
introduction of a non-vanishing local preference generates a
deterministic force, which can be described as an effective potential
$V(\delta)$ for the relative difference of densities
$\delta=(N_A-N_B)/N$. This potential has a minimum at the coexistence
state, $\delta=0$, corresponding to a maximum in the probability
distribution at $N_A=N_B=N/2$.  In other words, species coexistence
emerges for infinitely large sizes.  On the other hand, for finite
systems, when fluctuations are considered, the only possible true
steady states are the absorbing states $\delta=\pm 1$, where the
effective potential $V(\delta)$ is singular. The minimum at $\delta$
is separated from the negative singularities by two potential
barriers. As strength of the local preference and/or $N$ increase, the
basin of attraction of the coexistence state becomes larger and deeper
and the two symmetric barriers become closer to the absorbing states
and higher. Consequently the time needed to escape the coexistence
state becomes much longer, therefore unaccessible in computer
simulations.  Thus, three different regimes can thus be identified:
the absorbing, intermediate (quasi-active) and active phase (much as
in Fig. \ref{fig:nearneutral_coexistence}).  In the absorbing phase,
symmetry is broken and one of the two species reaches extinction with
certainty. This regime is equivalent to the monodominance regime in
Fig. \ref{fig:nearneutral_coexistence}.  The active phase is
characterized by a coexistence of both species, and survives
fluctuations only in the infinite-size limit. This corresponds to the
coexistence phase of Fig. \ref{fig:nearneutral_coexistence}.  Finally,
the intermediate state is a mixture of the two previous ones: the
absorbing states and the coexistence state are locally stable, thus,
the system is tri-stable, and the steady state depends on the initial
conditions. This is the mixed state of
Fig.~\ref{fig:nearneutral_coexistence}. These results provide a nice
analytical example of how noise can effectively change the shape of a
deterministic potential. Still, the presence of absorbing states --
with the associated singularities in the steady state distribution --
prevent true phase transitions from occurring: the only possible
steady state for any finite system is an absorbing one. Only in the
infinite-size limit, noise vanishes and the coexistence state becomes
truly stable \cite{Borile-JSTAT1}.

Another study scrutinized the case in which there are local
  habitat preference only at some specific locations in space, while
  all other sites are neutral \cite{Borile-JSTAT2}. An interesting
example which has been analyzed in details is that of a square lattice
where only the left (resp. right) boundary has a preference for
species $A$ (resp. $B$), (\cite{Borile-JSTAT2}, see also
\cite{Mobilia03,Mobilia07}).  The conclusion is that even mild biases
at a small fraction of locations induce robust and durable species
coexistence, also in regions arbitrarily far apart from the biased
locations. As carefully discussed in \cite{Borile-JSTAT2} this result
stems from the long-range nature of the underlying critical bulk
dynamics of the neutral voter model, and is robust to the introduction
of non-symmetrical biases --i.e. stronger for one of the species--
except for the fact that the state of coexistence is no longer
symmetric. These conclusions have a number of potentially important
consequences, for example, in conservation ecology as it suggests that
constructing local ``sanctuaries'' for different competing species can
result in global increase of stability of their populations, and
thus an enhancement of biodiversity, even in regions
arbitrarily distant from the protected zones \cite{Borile-JSTAT2}.

\subsection{Temporally-dependent  habitat preferences}

We have seen that spatial quenched disorder generically fosters
species coexistence. Another important question is what happens when
the preference for a species are time-dependent, i.e. if neutrality is
temporarily broken in favor of one of the coexisting species, while
the ecosystem remains neutral on average. This question has a long
tradition in ecology.  Several theoretical studies have looked at the
impact of environmental fluctuations on population growth and
ecosystem stability \cite{Chesson1981,Ridolfi}.  On one hand,
environmental stochasticity enhances fluctuations and extinction
rates, that can have a destabilizing effect on the ecological
community. On the other hand, it can also foster stability, as the
temporal alternance of species can effectively reduce the strength of
interspecific competition.

Similarly to the case of spatial disorder, one can design
quasi-neutral models where habitat-preferences for different species
are time-dependent, i.e. where in each time window there is a
preference for a randomly chosen species.  Different works have
recently analyzed this type of models, showing that time-dependent
habitat preference greatly improves predictions of empirical
ecological patterns with respect to purely neutral theories
\cite{Danino1,Danino2,Jordi,Hidalgo}. In particular, it has been
claimed that these models provides more realistic estimates of
dynamical quantities, such as average species persistence times and
distributions of species turnover \cite{azaele2006dynamical}, compared
with their neutral counterparts.

\subsection{Models with density dependence}

In ecology, one speaks of {\em density-dependence} or Allee effect
when the fitness of an individual depends on the abundance of the
species it belongs to. The underlying mechanisms can be very diverse,
from cooperative defense/feeding to spreading of parasites among
conspecific. An interesting scenario is that of negative
density-dependence, i.e. when individuals belonging to more abundant
species have lower fitness.  It is established that, in well mixed
systems, negative density-dependence significantly favors species
coexistence \cite{Chesson00}.  Versions of the voter model
implementing a negative density-dependence have been studied in the
literature \cite{Molofsky99,Schweitzer09}. In these models, the
reproduction probability of an individual depends on the number of
conspecific individuals in a given local neighborhood. Strictly
speaking, these models are not neutral: the neutral hypothesis is
defined at the level of individuals \cite{Hubbell2001}, and here
individuals belonging to species of different abundance do not have
the same fitness.  However, these models, as the other models
considered in this Section, are still symmetric, since all species are
treated on equal footing. Interesting phenomena like the possibility
of spontaneous breakdown of such a symmetry --thus leading to
asymmetric species coexistence-- have been recently uncovered at the
mean field level \cite{Borile-PRL}.

\section{Perspectives and conclusions\label{sec:conclusions}}

The range of ecological problems discussed in this review is by
no means exhaustive, and we believe there are many directions that
still need to be explored or fully understood.

A prominent example is the role of different speciation mechanisms on
spatial biodiversity. In the models discussed in this review,
speciation events involve a single individual ({\em point speciation
  mode}, in the language of evolutionary ecology). This assumption is
convenient from the modeling perspective, but leads to fitted values
of the speciation rate that are incompatible with independent
estimates \cite{ricklefs2006unified}. This assumption also tends to
generate too many young species which last for a short time and
overweights rare species. To address these issues, recently, another
mechanism called {\em protracted speciation} has been proposed in the
context of neutral models \cite{rosindell2010protracted}. In
protracted speciation, the speciation event does not occur at a single
generation, but is a gradual event lasting for some generations.
Introducing protracted speciation partially solves some of the
aforementioned problems \cite{rosindell2010protracted}. In real
ecosystems, even more speciation mechanisms are at play
\cite{speciation}.  For example, in {\em parapatric speciation}, two
spatially-separated population of the same species can diverge and
give rise to two different species. This would correspond to a
speciation event involving a group of individuals
rather than a single one. The role of different speciation modes in
maintaining biodiversity and in patterning the spatial organization of
species is still under discussion and modeling results can provide
very useful contributions to this debate.

As mentioned in the Introduction, ecological neutral theory elicited a
heated debate which is far from being solved as, in many cases,
non-neutral models based on the concept of niche and neutral models
yield similar fits of biodiversity patterns
\cite{chave2002comparing,McGill2003,tilman2004}. In recent years a new
view on this debate has been emerging. In Chase and Leibold's words:
``niche and neutral models are in reality two ends of a continuum with
the truth most likely in the middle''
\cite{chase2003ecological}. Indeed, the ecological forces underlying
niche and neutral models are not mutually exclusive, and demographic
stochasticity plays an important role also in non-neutral
settings. However, it has been difficult to clarify the importance of
different neutral and non-neutral mechanisms, as most non-neutral
model are characterized by a large number of parameters. Some progress
in this direction has been obtained in simplified settings which,
similarly to the model presented in Sect.~\ref{sec:nearneutral}, allow
for a controlled departure from neutrality. For instance, Haegeman and
Loreau \cite{Haegeman2011} added the main ingredients of neutral
theory, demographic stochasticity and immigration, to a Lotka-Volterra
competition model.  Similar problems have been studied in
Refs. \cite{Noble2011a,Noble2011b,Pigolotti2013}. An interesting
future direction would be to study similar models in a spatial
context.

In many ecological communities, in particular of microbial organisms,
ecological and evolutionary timescales are not separated.
Eco-evolutionary models describing both processes are becoming more
and more important \cite{Paula}. Neutral theory has provided a simple
framework to describe patterns in these communities, for example in
gut microbiota \cite{jeraldo2012quantification}. These systems call
for new theoretical efforts and new observables, such as
generalizations of the $\beta$-diversity taking into account genetic
differences among individuals \cite{houchmandzadeh2017neutral}.

We have seen throughout this review how some observables measured by
ecologists corresponds to well known quantities in statistical
physics: for instance, the $\beta$-diversity is closely related with a
two-point correlation function.  Other observables, such as SARs and
SADs, are less common in statistical physics.  A potentially fruitful
future direction is to consider other observables which are common in
statistical mechanics, such as multi-point correlation functions, and
measure them in ecosystems. In this direction, it is very interesting
the study of species clustering in \cite{plotkin2002cluster} based on
the theory of continuum percolation.

In summary, we presented an overview of different
stochastic spatial models in population ecology. We have seen that
even very simple models are a source of challenging problems in
statistical physics. In particular, because of speciation, each
species is bound to extinction and is therefore ultimately
transient. This feature is in contrast with traditional classical spin
system defined on a lattice where, even when in out-of-equilibrium
conditions, the number of spin components is fixed from the beginning.
Further, ecosystems are typically two-dimensional and, due to the
underlying diffusive behavior, $D=2$ is the critical dimension for
these models.  We have shown that this fact often leads to logarithmic
corrections to scaling laws, which have been difficult to analyze both
analytically and numerically. Despite these difficulties, remarkable
progress has been made in recent years. We believe that
cross-fertilization between statistical physics and ecology will 
  be more and more important in the future to deepen our quantitative
understanding of how ecosystems are organized.

\section*{Appendix: General scaling relationships}

In this brief Appendix, we discuss general condition imposed on the
functions $f$ and $g$ by the properties of the function $\Psi$,
depending on the exponent $\Delta$, see eq.(\ref{pnas}),
eq. (\ref{Delta}) and \cite{PNAS-Zillio}.
Let us write the normalization condition for $P(n;A)$
\begin{equation}
\sum_n P(n;A) \approx  g(A) f(A)
  \int_{n_0/f(A)}^\Lambda dx ~x^{-\Delta} =1 .
\end{equation}
The infrared cutoff $\Lambda$ is related to the fact that the function
$\psi(x)$ is a power-law for small $x$ only and rapidly decays for
larger arguments, see e.g. Fig. \ref{fig:panel}.
The integral is singular for small $x$ and  $\Delta >1$ and thus
\begin{equation} 1 \sim g(A) f(A) f(A)^{\Delta-1} =
  g(A)f(A)^{\Delta}\ .
\end{equation}
On the other hand, if $\Delta<1$, the integral is weakly dependent on $f(A)$,
so that
\begin{equation} 
1 \sim g(A) f(A)\ .
\end{equation}
Similarly, the first moment of $\Psi$ is
  \begin{equation} \langle n \rangle \sim g(A) f^2(A) 
f(A)^{\Delta-1} = g(A) f(A)^{\Delta+1}
\end{equation}
if $1<\Delta <2$ and
\begin{equation}  \langle n \rangle  \sim g(A) f^2(A)
\end{equation}
for $ \Delta >2$.  Combining the expressions above,
different regimes emerge as a function of $\Delta$: if $\Delta <1$, $
f(A) = \langle n \rangle $, while for $1< \Delta <2$, $ f(A) = \langle
n \rangle^{1/(2-\Delta)} $, while no specific prediction for $f(A)$
can be made in the case $\Delta \geq 2$.  In particular, for $\Delta
<1$ one has a simple scaling form $f(A) =\langle n \rangle $ and
$g(A)= 1/\langle n \rangle $ which applies, for example, to the $1D$
case as described in the main text.  The marginal case $\Delta=1$ is
treated in detail in Sec. \ref{sec:sad}.

\begin{acknowledgements}
MAM is grateful to the Spanish-MINECO for financial
support (under grant FIS2013-43201-P; FEDER funds), as well as to
J. Hidalgo, S. Suweis, A. Maritan, C. Borile for a long term
collaboration on topics related to the content of this paper.
\end{acknowledgements}

\bibliography{neutral} 

\end{document}